\documentclass[12pt]{article}%
\usepackage{amssymb}
\usepackage{amsfonts}
\usepackage{amsmath}
\usepackage[round]{natbib}
\usepackage{makeidx}
\usepackage{mathtools}
\renewcommand{\cite}{\citet}
\usepackage[utf8]{inputenc}
\usepackage[T1]{fontenc}
\usepackage[left=2cm,right=2cm,top=2cm,bottom=2cm]{geometry}

\usepackage[singlespacing]{setspace}
\usepackage[bottom]{footmisc}
\usepackage{indentfirst}
\usepackage{endnotes}
\usepackage{graphicx}
\usepackage{epstopdf}
\usepackage{rotating}
\usepackage{verbatim}
\usepackage{setspace}
\usepackage{multirow}
\usepackage{latexsym}
\usepackage{bigints}
\usepackage{proba}
\allowdisplaybreaks

\usepackage{amsthm}
\usepackage{makeidx}
\usepackage{fancyhdr}
\usepackage{type1cm}
\usepackage{tikz}
\usepackage{subcaption}

\usepackage{booktabs}
\DeclareGraphicsExtensions{.pdf,.png,.jpg}

\newcommand{\wh}{\widehat}

\newcommand{\wt}{\widehat{\theta}}
\newcommand{\whp}{\widehat{\psi}}

\newcommand{\wtl}{\widetilde}
\newcommand{\lt}{\left}
\newcommand{\rt}{\right}
\newcommand{\EE}{\mathbf{E}}

\newtheorem{example}{Example}

\newcommand{\arginf}{\mathrm{arginf}}

\newcommand{\beq}{\begin{eqnarray*}}
\newcommand{\eeq}{\end{eqnarray*}}

\newtheorem{thm}{Theorem}[section]

\newtheorem{df}{Definition}[section]
\newtheorem{lem}{Lemma}[section]
\newtheorem{cor}{Corollary}[section]

\newtheorem{assum}{Assumption}

\numberwithin{equation}{section}

\theoremstyle{definition}

\usepackage[pdfpagemode=UseOutlines ,plainpages=false
,hypertexnames=false ,pdfpagelabels ,hyperindex=true,colorlinks=true]{hyperref}
        \makeatletter%
        \Hy@breaklinkstrue%
        \makeatother%
\usepackage{color}
\definecolor{darkred}{rgb}{0.6,0.0,0.1}
\definecolor{darkgreen}{rgb}{0,0.5,0}
\definecolor{darkblue}{rgb}{0,0,0.5}
\hypersetup{colorlinks ,linkcolor=darkblue ,filecolor=darkgreen
,urlcolor=darkblue ,citecolor=black}

\definecolor{dgreen}{rgb}{0,0.5,0}
\definecolor{dblue}{rgb}{0,0,0.9}
\definecolor{dred}{rgb}{0.6,0.0,0.1}
\definecolor{dgold}{rgb}{0.5,0.3,0.0}
\definecolor{dvio}{rgb}{0.6,0.3,0.5}
\definecolor{gray}{rgb}{0.5,0.5,0.5}

\newcommand{\bee}{\begin{equation}}
\newcommand{\eee}{\end{equation}}

\makeatletter
\def\@biblabel#1{\hspace*{-\labelsep}}
\makeatother
\usepackage{xr}
\begin{document}

\title{\vspace{-1in}Bayesian Estimation and Comparison of Moment Condition Models\thanks{The authors gratefully thank the Co-Editor, an Associate Editor, and four anonymous referees for their many constructive comments on the previous version of the paper.}}
\author{{\sc Siddhartha Chib}\thanks{Olin Business School, Washington University in St. Louis, Campus Box 1133, 1 Brookings Dr. St. Louis, MO 63130, USA, e-mail:
\url{chib@wustl.edu}}\\
{\small \it  Washington University in St. Louis}
  \and{\sc Minchul Shin}\thanks{Department of Economics, University of Illinois, 214 David Kinley Hall, 1407 W. Gregory, Urbana, IL 61801, e-mail: \url{mincshin@illinois.edu}}\\
  {\small \it University of Illinois}
  \and  {\sc Anna Simoni}\thanks{CREST - ENSAE - \'{E}cole Polytechnique, 15, Boulevard Gabriel P\'{e}ri, 92240 Malakoff, France, e-mail: \url{simoni.anna@gmail.com}}\\
  {\small \it CREST, CNRS}
  }

\date{This version: \today \\ First version: June, 2016}
\maketitle
\begin{abstract}
  In this paper we consider the problem of inference in statistical models characterized by moment restrictions
  by casting the problem within the Exponentially Tilted Empirical Likelihood (ETEL) framework. Because the ETEL function has a well defined probabilistic interpretation and plays the role of a nonparametric likelihood, a fully Bayesian semiparametric framework can be developed. We establish a number of powerful results surrounding the Bayesian ETEL framework in such models. One major concern driving our work is the possibility of misspecification. To accommodate this possibility, we show how the moment conditions can be reexpressed in terms of additional nuisance parameters and that, even under misspecification, the Bayesian ETEL posterior distribution satisfies a Bernstein-von Mises  result. A second key contribution of the paper is the development of a framework based on marginal likelihoods and Bayes factors to compare models defined by different moment conditions. Computation of the marginal likelihoods is by the method of \cite{Chib1995} as extended to Metropolis-Hastings samplers in \cite{ChibJeliazkov2001}. We establish the model selection consistency of the marginal likelihood and show that the marginal likelihood favors the model with the minimum number of parameters and the maximum number of valid moment restrictions. When the models are misspecified, the marginal likelihood model selection procedure selects the model that is closer to the (unknown) true data generating process in terms of the Kullback-Leibler divergence. The ideas and results in this paper provide a further broadening of the theoretical underpinning and value of the Bayesian ETEL framework with likely far-reaching practical consequences. The discussion is illuminated through several examples.
\end{abstract}


\setlength{\baselineskip}{21.0pt}

\section{Introduction}

Our goal in this paper is to develop a Bayesian analysis of moment condition models. By moment condition models, we mean models that are specified only through moment restrictions of the type $\EE^{P}[g(X,\theta )]=0$, where $g(X,\theta )$ is a known vector-valued function of a random vector $X$ and an unknown parameter vector $\theta $, and $P$ is the unknown data distribution. Models of this type, which arise frequently in statistics and econometrics, see \textit{e.g.} \cite{BroniatowskiKeziou2012}, can be attractive since full modeling of $P$ is not invoked and inferences about $\theta$ are based only on the partial information supplied by the set of moment conditions. For instance, in a regression context, letting $X=(y,x)$ and $y=x\beta +\varepsilon $, where $y$ is the scalar response and $x$ is a scalar predictor, one can learn about the regression parameter $\beta $ from the orthogonality assumption $\EE^{P}[(y-x\beta )x]=0$ without fully modeling the error distribution or the parameters of the error distribution. More generally, $\beta$ can be inferred in this setting from the orthogonality conditions $\EE^{P}[(y-x\beta )z] = 0$, given a set of instrumental variables $z$. Examples of such moment condition models abound, but for the most part the analysis of such models from the Bayesian perspective has proved elusive since typical parametric and semiparametric Bayesian methods are reliant on a full probability model of $P$.

On the frequentist side, the recent developments in empirical likelihood (EL) based methods, see \textit{e.g.}  \cite{Owen1988,owen1990,Owen2001}, \cite{QinLawless1994}, \cite{KitamuraStutzer1997}, \cite{Imbens1997}, \cite{Schennach2007}, \cite{ChenVanKeilegom2009}, and references therein, have opened up a promising approach for dealing with moment condition models. There are emerging cogent arguments for using the EL in Bayesian analysis. For example, \cite{Lazar2003} has argued that the EL can be used in a Bayesian framework in place of the data distribution $P$. In fact, \cite{Schennach2005} shows that it is possible to obtain a nonparametric likelihood closely related to EL, called the exponentially tilted empirical likelihood (ETEL), by marginalizing over $P$ with a nonparametric prior that favors distributions that are close to the empirical distribution function in terms of the Kullback-Leibler (KL) divergence while satisfying the moment restrictions. In addition, \cite{grendar2009} show that the EL is the mode of the posterior of $P$ under a general prior on $P$. Thus, by combining either the EL or the ETEL functions with a prior $\pi(\theta)$ on $\theta $, moment condition models can in principle be subjected to a Bayesian semiparametric analysis. Applications of this idea are given for instance by \cite{LancasterJun2010}, \cite{KimYang2011}, \cite{YangHe2012}, \cite{xi2016} to handle moment condition models, by \cite{rao2010} in complex survey estimation, and by \cite{chaudhuri2011}, \cite{porter2015}, \cite{Chaudhuri2017} in small area estimation. On the theory side, \cite{YangHe2012} shows the asymptotic normality of the Bayesian EL posterior distribution of the quantile regression parameter, and \cite{fang2006} and \cite{chang2008} study the higher-order asymptotic and coverage properties of the Bayesian EL/ETEL posterior distribution for the population mean, while \cite{Schennach2005} and \cite{LancasterJun2010} consider the large-sample behavior of the Bayesian ETEL posterior distribution under the assumption that all moment restrictions are valid. Alternative, non-EL/ETEL based approaches for moment condition models, which we do not consider in this paper, have also been examined, for example, \cite{BornnShephardSolgi2015}, \cite{FlorensSimoni2015} and \cite{KitamuraOtsu2011}.

The goal of this paper is to establish a number of powerful results for the Bayesian analysis of moment condition models, within the ETEL framework, complementing and extending the aforementioned papers in important directions. One major goal is the Bayesian analysis of moment condition models that are potentially misspecified. For this reason, our analysis is built on the ETEL function which, as shown by \cite{Schennach2007}, leads to frequentist estimators of $\theta$ that have the same orders of bias and variance (as a function of the sample size) as the EL estimators but, importantly, maintain the root $n$ convergence even under model misspecification (see \cite[Theorem 1]{Schennach2007}). We show that the ETEL framework is an immensely useful organizing framework within which a fully Bayesian treatment of correctly and misspecified moment condition models can be developed. We show that even under misspecification, the Bayesian ETEL posterior distribution has desirable properties, and that it satisfies the Bernstein - von Mises (BvM) theorem.

Another key focus of the paper is the development of a unified framework based on marginal likelihoods and Bayes factors for comparing different moment restricted models and for discarding any misspecified moment restrictions. For an overview on Bayesian model selection in standard models we refer to \cite{Robert2002} and references therein. Essentially, each set of moment restrictions, and the different sets of restrictions on the parameters, define different models. Our proposal is to select the model with the largest marginal likelihood. It may be noted that since one aim of our model selection comparison is to discard misspecified moment restrictions, we do not consider the model averaging perspective. In order to operationalize model comparisons in our set-up, in particular when models are defined by different numbers of moment conditions, it is necessary to linearly transform the moment functions $g(X,\theta)$ so that all the transformed moments are included in each model. This linear transformation simply consists of adding an extra parameter different from zero to the components of the vector $g(X,\theta)$ that correspond to the restrictions not included in a specific model. This procedure is explained in detail below.

We compute the marginal likelihood by the method of \cite{Chib1995}, as extended to Metropolis-Hastings samplers in \cite{ChibJeliazkov2001}. This method renders computation of the marginal likelihood simple and is a key feature of both our numerical and theoretical analysis. Our formal asymptotic theory covers the following exhaustive possibilities: the case where the models in the comparison set contain only valid moment restrictions, the case where all the models in the set are misspecified, and finally the case where some of the models contain only valid moment restrictions while the others contain at least one invalid moment restriction. Our analysis shows that the marginal likelihood based selection procedure is consistent in the sense that: \textit{(i)} it discards misspecified moment restrictions, \textit{(ii)} it selects the model that is the ``less misspecified'' when comparing models that are all misspecified, \textit{(iii)} it selects the model that contains the maximum number of overidentifying valid moment restrictions when comparing correctly specified models, and \textit{(iv)} when some models are correctly specified and some are misspecified, it selects the model that is correctly specified and contains the maximum number of overidentifying moment conditions. These fundamental model selection consistency results are based on the asymptotic behavior of the ETEL function, and the validity of the BvM theorem, both under correct specification and misspecificition. These results, developed within an entirely formal Bayesian setting, can be viewed as complementary to the less Bayesian formulations described in \cite{Variyath2010} and \cite{vexler2013} where the focus is on quasi-Bayes factors constructed from the EL, and \cite{HongPreston2012} where models are compared based on a quasi-marginal likelihood obtained from an approximation to the true $P$. In brief, the strategy for comparing models developed in this paper, along with the associated large sample theory and striking results, should have far-reaching practical consequences.

The rest of the article is organized as follows. In Section \ref{s_The_setting} we describe the moment condition model, define the notion of misspecification in this setting, and then discuss the prior-posterior analysis with the ETEL function. We then provide the first pair of major results dealing with the asymptotic behavior of the posterior distribution for both correctly specified and misspecified models. Section \ref{s_BMS} introduces our model selection procedure based on marginal likelihoods and the associated large sample results. Throughout the paper, for expository purposes, we include numerical examples. Then in Section \ref{s_examples} we discuss the problems of variable selection in a count regression model and instrument validity in an instrumental variable regression. Section \ref{s_conclusion} concludes. Proofs of our results are collected in the Appendix and in a Supplementary Appendix.

\section{Setting}

\label{s_The_setting}

Suppose that $X$ is an $\mathbb{R}^{d_{x}}$-valued random vector with
(unknown) distribution $P$. Suppose that the operating assumption is that
the distribution $P$ satisfies the $d$ unconditional moment restrictions
\begin{equation}
\EE^{P}[g(X,\theta )]=0  \label{eq_moment_restrictions}
\end{equation}%
\noindent where $\EE^{P}$ denotes the expectation taken with respect to $P$, $g:\mathbb{R}^{d_{x}}\times \Theta \mapsto \mathbb{R}^{d}
$ is a vector of known functions with values in $\mathbb{R}^{d}$, $\theta
:=(\theta _{1},\ldots ,\theta _{p})^{\prime }\in \Theta \subset \mathbb{R}%
^{p}$ is the parameter vector of interest, and $0$ is the $d\times 1$ vector of zeros.
We assume that $\EE^{P}[g(X,\theta )]$ is bounded for every $\theta \in
\Theta $. We also suppose that we are given a random sample $x_{1:n}:=(x_{1},\ldots ,x_{n})$ on $X$ and that $d\geq p$.

When the number of moment restrictions $d$ exceeds the number of parameters $p$, the parameter $\theta $ in such a setting is said to be overidentified (over restricted). In such a case, there is a possibility that a subset of the moment condition may be invalid in the sense that the true data generating process is not contained in the collection of probability measures that satisfy the moment conditions for all $\theta \in \Theta $. That is, there is no parameter $\theta $ in $\Theta$ that is consistent with the moment restrictions \eqref{eq_moment_restrictions} under the true data generating process $P$. To deal with possibly invalid moment restrictions, we reformulate the moment conditions in terms of an additional nuisance parameter $V\in \mathfrak{V}\subset \mathbb{R}^{d}$. For example, if the $k$-th moment condition is not expected to be valid, we subtract $V = \left(V_{1},\ldots,V_{d}\right)$ from the moment restrictions where $V_{k}$ is a free parameter and all other elements of $V$ are zero. To accommodate this situation, we rewrite the above conditions as the following augmented moment conditions
\begin{equation}
\EE^{P}[g^{A}(X,\theta,V)]=0  \label{eq_augmented_model}
\end{equation}%
\noindent where $g^{A}(X,\theta ,V):=g(X,\theta )-V$. Note that in this formalism, the parameter $V$ indicates which moment restrictions are active where for `active moment restrictions' we mean the restrictions for which the corresponding components of $V$ is zero. In order to guarantee identification of $\theta$, at most $(d-p)$ elements of $V$ can be different than zero. If all the elements of $V$ are zero, we recover the restrictions in \eqref{eq_moment_restrictions}.

Let $d_v \leq (d-p)$ be the number of non-zero elements of $V$ and let $v\in\mathcal{V}\subset \mathbb{R}^{d_v}$ be the vector that collects all the non-zero components of $V$. We call $v$ the augmented parameter and $\theta$ the parameter of interest. Therefore, the number of active moment restrictions is $d - d_v$. In the following, we write $g^A(X,\theta,v)$ as a shorthand for $g^A(X,\theta,V)$, with $v$ the vector obtained from $V$ by collecting only its non-zero components.

The central problem of misspecification of the moment conditions, mentioned
in the preceding paragraph, can now be formally defined in terms of the
augmented moment conditions.

\begin{df}[Misspecified model]\label{def_misspecified_model}
  We say that the augmented moment condition model is misspecified if the set of probability measures implied by the moment restrictions does not contain the true data generating process $P$ for every $(\theta,v)\in\ \Theta \times \mathcal{V}$, that is, $P\notin \mathcal{P}$ where $\mathcal{P} = \bigcup_{(\theta,v)\in\Theta \times \mathcal{V}}\mathcal{P}_{(\theta,v)}$ and $\mathcal{P}_{(\theta,v)} = \{Q\in \mathbb{M}; \,\EE^Q[g^{A}(X,\theta,v)] = 0\}$ with $\mathbb{M}$ the set of all probability measures on $\mathbb{R}^{d_x}$.
\end{df}

In a nutshell, a set of augmented moment conditions is misspecified if there is no pair $(\theta ,v)$ in $(\Theta \times \mathcal{V})$ that satisfies $\EE^{P}[g^{A}(X,\theta ,v)]=0$ where $P$ is the true data generating process. On the other hand, if such a pair of values $(\theta,v)$ exists then the set of augmented moment conditions is correctly specified.

Throughout the paper, we use regression models to understand the various concepts and ideas.

\begin{example}[Linear regression model]
  Suppose that we are interested in estimating the following linear regression model with an intercept and a predictor:
\bee \label{eq:ex_reg}
y_{i} = \alpha + \beta z_{i} + e_{i},\qquad i=1,\ldots,n
\eee
where $(z_{i}, e_{i})'$ are independently drawn from some distribution $P$. %for $i=1,2,...,n$.
Under the assumption that $\EE^{P}[e_{i}|z_{i}]=0$, we can use the following moment restrictions to estimate $\theta := (\alpha, \beta)$:
\bee \label{eq:ex_reg_mom}
\EE^{P}[e_{i}(\theta) ] = 0, \quad \EE^{P}[e_{i}(\theta)  z_{i}]=0, \quad \EE^{P}[(e_{i}(\theta))^{3}] = v,
\eee
where $e_{i}(\theta) := (y_{i} - \alpha - \beta z_{i})$. The first two moment restrictions are derived from the standard orthogonality condition and identify $\theta$. The last restriction potentially serves as additional information.
In terms of the notation in \eqref{eq_moment_restrictions} and \eqref{eq_augmented_model}, $x_i:=(y_i,z_{i})$, $g(x_i,\theta) = (e_i(\theta),e_i(\theta)z_{i},e_i(\theta)^3)'$, $V=(0,0,v)'$, $d_v = 1$ and $g^A(x_i,\theta,V) = g(x_i,\theta) - (0,0,v)'$. If one believes that the underlying distribution of $e_{i}$ is indeed symmetric, then one could use this information by setting $v$ to zero. Otherwise, it is desirable to treat $v$ as an unknown object. If the distribution of $e_{i}$ is skewed and $v$ is forced to be zero, then the model becomes misspecified because no $(\alpha, \beta)$ can be consistent with the three moment restrictions jointly under $P$. When the augmented parameter $v$ is treated as a free parameter, the model is correctly specified even under asymmetry.
\end{example}

%=====================================================================
\subsection{Prior-Posterior analysis}

\label{ss_prior_posterior}

Consider now the question of prior-posterior analysis under the ETEL function. Although our setting is similar to that of \cite{Schennach2005}, the presence of the augmented parameter $v$ and the possibility of misspecification, lead to a new analysis and new results.

For any $(\theta,v)$, define the convex hull of $\bigcup_{i=1}^{n}g^{A}(x_{i},\theta ,v)$ as the following convex subset of $\mathbb{R}^d$: $\{\sum_{i=1}^n p_i g^A(x_i,\theta,v); p_i\geq 0, \forall i=1,\ldots,n, \, \sum_{i=1}^n p_i = 1\}$.
Now suppose that \textit{(i)} $g^{A}(x,\theta ,v)$ is continuous in $x$ for every $(\theta ,v)\in \Theta \times \mathcal{V}$ (or has a finite number of step discontinuities) and \textit{(ii)} the interior of the convex hull of $\bigcup_{i=1}^{n}g^{A}(x_{i},\theta ,v)$ contains the origin. Suppose also that the nonparametric prior on $P$ is the mixture of uniform probability densities described in \cite{Schennach2005}, which is capable to approximating any distribution as the number of mixing components increases. Then, adapting the arguments of \cite{Schennach2005}, the posterior distribution of $(\theta ,v)$ after marginalization over $P$ has the form
\begin{equation}
\pi (\theta ,v|x_{1:n})\propto \pi (\theta ,v)p(x_{1:n}|\theta ,v)
\label{eq: betel post}
\end{equation}%
where $\pi (\theta ,v)$ is the prior of $(\theta,v)$ and $p(x_{1:n}|\theta ,v)$
is the ETEL function defined as%
\begin{equation}\label{eq_ETEL_function}
p(x_{1:n}|\theta ,v)=\prod_{i=1}^{n}p_{i}^{\ast }(\theta ,v)
\end{equation}%
and $p_{i}^{\ast }(\theta ,v)$ are the probabilities that minimize the
KL divergence between the probabilities $(p_{1},\ldots,p_{n})$ assigned to each sample observation and the empirical probabilities $(\frac{1}{n},\ldots,\frac{1}{n})$, subject to the conditions that the probabilities $(p_{1},\ldots,p_{n})$ sum to one and that the expectation under these probabilities satisfies the given moment conditions:%
\begin{eqnarray}
&&\max_{p_{1},\ldots ,p_{n}}\sum_{i=1}^{n}\left[ -p_{i}\log (np_{i})\right]
\label{eq:Entropy_0}\\
\text{subject to}\qquad \sum_{i=1}^{n}p_{i} &=&1\quad \text{and }\quad
\sum_{i=1}^{n}p_{i}g^{A}(x_{i},\theta ,v)=0.  \label{eq: Entropy}
\end{eqnarray}

For numerical and theoretical purposes below, the preceding probabilities are computed more conveniently
from the dual (saddlepoint) representation as, for $i=1,\ldots,n$
\begin{equation}\label{eq_ETEL_weights}
p_{i}^{\ast }(\theta ,v):=\frac{e^{\wh{\lambda}(\theta ,v)^{\prime
}g^{A}(x_{i},\theta ,v)}}{\sum_{j=1}^{n}e^{\wh{\lambda}(\theta ,v)^{\prime
}g^{A}(x_{j},\theta ,v)}},\quad \text{ where }\wh{\lambda}(\theta ,v)=\arg
\min_{\lambda \in \mathbb{R}^{d}}\frac{1}{n}\sum_{i=1}^{n}\exp \left(
\lambda ^{\prime }g^{A}(x_{i},\theta ,v)\right) .
\end{equation}%
Therefore, the posterior distribution takes the form
\begin{equation}
\pi (\theta ,v|x_{1:n})\propto \pi (\theta ,v)\prod_{i=1}^{n}\frac{e^{\wh{\lambda}(\theta ,v)^{\prime }g^{A}(x_{i},\theta ,v)}}{\sum_{j=1}^{n}e^{\wh{\lambda}(\theta ,v)^{\prime }g^{A}(x_{j},\theta ,v)}},  \label{eq:BETELpost}
\end{equation}%
which may be called the Bayesian Exponentially Tilted Empirical Likelihood (BETEL) posterior distribution. It can be efficiently simulated by Markov chain Monte Carlo (MCMC) methods. For example, the one block tailored Metropolis-Hastings (M-H) algorithm \citep{ChibGreenberg1995} is applied as follows. Let $q(\theta ,v|x_{1:n})$ denote a student-t distribution whose location parameter is the mode of the log ETEL function and whose dispersion matrix is the inverse of the negative Hessian matrix of the log ETEL function at the mode. Then, starting from some initial value $(\theta^{(0)},v^{(0)})$, we get a sample of draws from the BETEL posterior by repeating the following steps for $s=1,\ldots,S$:

\begin{enumerate}
\item Propose $(\theta ^{\dagger },v^{\dagger })$ from $q(\theta ,v|x_{1:n})$ and solve for $p_{i}^{\ast }(\theta ^{\dagger },v^{\dagger })$, $1\leq i\leq
n$, from the Exponential Tilting saddlepoint problem \eqref{eq_ETEL_weights}.

\item Calculate the M-H probability of move
\begin{equation*}
\alpha \left(\left.(\theta ^{s-1},v^{s-1}),(\theta ^{\dagger },v^{\dagger})\right|x_{1:n}\right)=\min \left\{ 1,\frac{\pi (\theta ^{\dagger },v^{\dagger
}|x_{1:n})}{\pi (\theta ^{s-1},v^{s-1}|x_{1:n})}\frac{q(\theta
^{s-1},v^{s-1}|x_{1:n})}{q(\theta ^{\dagger },v^{\dagger }|x_{1:n})}\right\}.
\end{equation*}

\item Set $(\theta ^{s},v^{s})=(\theta ^{\dagger },v^{\dagger })$ with
probability $\alpha ((\theta ^{s-1},v^{s-1}),(\theta ^{\dagger },v^{\dagger
})|x_{1:n})$. Otherwise, set $(\theta ^{s},v^{s})=(\theta ^{s-1},v^{s-1})$.
Go to step 1.
\end{enumerate}

Note that when the dimension of $(\theta ,v)$ is large, the Tailored
Randomized Block M-H algorithm of \cite{ChibRamamurthy2010}
can be used instead for improved simulation efficiency.

\paragraph{Prior specification.} In our examples, we focus on two prior distributions. Under the first prior, which we call the default prior, each element $\theta_k$ and $v_l$ of $\theta$ and $v$, respectively, is
given independent student-t distributions with $\nu = 2.5$ degrees of freedom, location zero and dispersion equal to 5:
\bee
\theta_k \sim t_{2.5}(0,5^2) \quad \text{ and } \quad v_l \sim t_{2.5}(0,5^2).
\label{eq:defaultprior}
\eee
In the second prior, which we call the training sample prior, an initial portion of the sample (which is not used for subsequent inferences) is used to find the ETEL estimate of the unknown parameters, that is, the maximizer of the ETEL function \eqref{eq_ETEL_function} whose definition is recalled in \eqref{eq_ETEL_def} in the Appendix. Then, the prior of each element of $(\theta',v')'$ is equal to the default prior except that now the location is set equal to the corresponding ETEL estimate.

To see the different implications of these prior distributions, consider two moment condition models defined by the  restrictions:
\bee
\begin{split}
M_{1} &: \quad \EE^{P}[g_{1}(X, \beta)] = 0, \quad \EE^{P}[g_{2}(X, \beta)] = 0 \\
M_{2} &: \quad \EE^{P}[g_{1}(X, \beta)] = 0, \quad \EE^{P}[g_{2}(X, \beta)] = v
\end{split}
\eee
where both moments restrictions are active under $M_{1}$ but only the first is active under $M_{2}$. Then, under
the default prior, a prior mean of 0 on $v$ implies the belief that the second moment restriction is likely to hold.
On the other hand, in the training sample prior, the prior location of $v$ is determined by the ETEL estimate of $v$ in the training sample. If this is substantially different from zero (relative to the prior dispersion) this prior implies the belief that the second moment restriction is, a priori, less likely to be active.

\setcounter{example}{0}
\begin{example}[continued]
  To illustrate the prior-posterior analysis, we generate $y_{i}$, $i=1,\ldots,n$  from the regression model in \eqref{eq:ex_reg} with the covariate $z_i\sim \mathcal{N}(0.5,1)$, intercept $\alpha = 0$, slope $\beta = 1$ and $e_{i}$ distributed according to the skewed distribution:
  \bee
    e_{i} \sim
    \begin{cases}
    \mathcal{N}(0.75,  0.75^{2}) \quad \text{with probability } 0.5 \\
    \mathcal{N}(-0.75, 1.25^{2}) \quad \text{with probability } 0.5.
    \end{cases}
    \label{eq:dgp3}
  \eee
  Our analysis is based on the moment restrictions in \eqref{eq:ex_reg_mom}, that is,
  \[
  g^A(x_i,\theta,v) = (e_i(\theta),e_i(\theta)z_i,e_i(\theta)^3 - v)',\qquad e_i(\theta) = y_i - \alpha - \beta z_i,
  \]
  with $\theta = (\alpha,\beta)$. These moment conditions are correctly specified because $v$ is free. Under the default independent student-t prior in (\ref{eq:defaultprior}), the marginal posterior distributions of $\alpha$, $\beta$ and $v$ are summarized in Table \ref{table:example1} for two different values of $n$. It can be seen from the .025 and .975 quantiles (called ``lower'' and ``upper'', respectively) that the marginal posterior distributions of $\alpha$ and $\beta$ are already concentrated around the true values for $n=250$ but concentrate even more closely around the true values for $n=2000$. This example showcases the ease with which such Bayesian inferences are possible.
  \begin{table}[ht]
\centering
\begin{tabular}{rrrrrrr}
  \hline
 & mean & sd & median & lower & upper & ineff \\ \hline
  & & &$n=250$ & & & \\ \hline
$\alpha$ & -0.03 & 0.10 & -0.03 & -0.24 & 0.16 & 1.49 \\
  $\beta$ & 0.99 & 0.08 & 0.98 & 0.83 & 1.15 & 1.70 \\
  $v$ & -1.42 & 0.36 & -1.39 & -2.20 & -0.82 & 3.21 \\ \hline
  & & & $ n = 2000$ & & & \\ \hline
  $\alpha$ & 0.00 & 0.03 & 0.00 & -0.06 & 0.06 & 1.30 \\
  $\beta$ & 0.99 & 0.03 & 0.99 & 0.93 & 1.04 & 1.21 \\
  $v$ & -0.97 & 0.10 & -0.97 & -1.18 & -0.78 & 1.34 \\ \hline
\end{tabular}
\caption{\small{Posterior summary for two simulated sample sizes from Example 1 (a regression model with skewed error distribution). The true value of $\alpha$ is $0$ and that of $\beta$ is $1$. The summaries are
based on $10,000$ MCMC draws beyond a burn-in of $1000$. The M-H acceptance rate is around $90\%$ in both cases. ``Lower'' and ``upper'' refer to the $.025$ and $.975$ quantiles of the simulated draws, respectively, and ``ineff'' to the inefficiency factor, the ratio of the numerical variance of
the mean to the variance of the mean assuming independent draws: an inefficiency
factor close to 1 indicates that the MCMC draws, although serially correlated, are essentially independent.}}
\label{table:example1}
\end{table}
\end{example}

\paragraph{Notation.} In Sections \ref{s_Asymptotic_result_cs} and \ref{s_Asymptotic_result_ms} we use the following notations. For ease of exposition, we denote $\psi := (\theta,v)$, $\psi\in\Psi$ with $\Psi := \Theta\times \mathcal{V}$. Moreover, $\|\cdot\|_F$ denotes the Frobenius norm and $\|\cdot \|$ the Euclidean norm. The notation `$\overset{p}{\to}$' is for convergence in probability with respect to the product measure $P^n = \bigotimes_{i=1}^n P$. The log-likelihood function for one observation is denoted by $l_{n,\psi}$:
$$l_{n,\psi}(x) := \log \frac{e^{\wh{\lambda}(\psi)'g^A(x,\psi)}}{\sum_{j=1}^n e^{\wh{\lambda}(\psi)'g^A(x_j,\psi)}}
= -\log n + \log \frac{e^{\wh \lambda'g^A(x,\psi)}}{\frac{1}{n}\sum_{j=1}^n\left[e^{\wh\lambda'g^A(x_j,\psi)}\right]}$$
so that the log-ETEL function is $\log p(x_{1:n}|\psi) = \sum_{i=1}^n l_{n,\psi}(x_i)$. For a set $\mathcal{A}\subset\mathbb{R}^m$, we denote by $int(\mathcal{A})$ its interior relative to $\mathbb{R}^m$. Further notations are introduced as required.

%===========================================================================================================================================
\subsection{Asymptotic Properties: correct specification}\label{s_Asymptotic_result_cs}
In this section, we first introduce additional notations and assumptions for correctly specified models. Under these assumptions, we establish both the large sample behavior of the BETEL posterior distribution and, in Section \ref{s_BMS}, the model selection consistency of our marginal likelihood procedure.

Let $\theta_*$ be the true value of the parameter of interest $\theta$ and $v_*$ be the true value of the augmented parameter. So, $\psi_*:=(\theta_*,v_*)$. The true value $v_*$ is equal to zero when the non-augmented model \eqref{eq_moment_restrictions} is correctly specified.
Moreover, let $\Delta := \EE^P[g^A(X,\psi_*)g^A(X,\psi_*)^{\prime }]$ and $\Gamma := \EE^P\left[\frac{\partial}{\partial \psi^{\prime }}g^A(X,\psi_*)\right]$. Assumption 1 requires that the augmented model is correctly specified in the sense that there is a value of $\psi$ such that \eqref{eq_augmented_model} is satisfied by $P$, and that this value is unique. A necessary condition for the latter is that $(d-p)\geq d_v \geq 0$.
\begin{assum}\label{Ass_0_NS}
  Model \eqref{eq_augmented_model} is such that $\psi_* \in \Psi$ is the unique solution to $\EE^P[g^A(X,\psi)] = 0$.
\end{assum}
The following two assumptions relate to the smoothness of the function $g^A(x,\psi)$, its moments, and the parameter space.
\begin{assum}\label{Ass_1_NS}
  (a) $X_i$, $i=1,\ldots, n$ are i.i.d. random variables that take values in $(\mathcal{X},\mathfrak{B}_{\mathcal{X}})$ with probability distribution $P$, where $\mathcal{X}\subseteq \mathbb{R}^{d_x}$; (b) for every $0 \leq d_v\leq d-p$, $\psi\in\Psi\subset \mathbb{R}^p\times \mathbb{R}^{d_{v}}$ where $\Theta$ and $\mathcal{V}$ are compact and connected and $\Psi := \Theta\times \mathcal{V}$; (c) $g(x,\theta)$ is continuous at each $\theta\in\Theta$ with probability one; (d) $\EE^P[\sup_{\psi\in\Psi}\|g^A(X,\psi)\|^{\alpha}] < \infty$ for some $\alpha > 2$; (e) $\Delta$ is nonsingular.
\end{assum}

\begin{assum}\label{Ass_2_NS}
  (a) $\psi_*\in int(\Psi)$; (b) $g^A(x,\psi)$ is continuously differentiable in a neighborhood $\mathfrak{U}$ of $\psi_*$ and $\EE^P[\sup_{\psi\in\mathfrak{U}}\|\partial g^A(X,\psi)/\partial \psi'\|_F]< \infty$; (c) $rank(\Gamma) = p$.
\end{assum}

\noindent Assumptions \ref{Ass_1_NS} and \ref{Ass_2_NS} are the same as the assumptions of \cite[Theorem 3.2]{NeweySmith2004} and \cite[Theorem 3]{Schennach2007}. The next assumption concerns the prior distribution and is
a standard assumption to establish asymptotic properties of Bayesian procedures.

\begin{assum}\label{Ass_prior}
  (a) $\pi$ is a continuous probability measure that admits a density with respect to the Lebesgue measure; (b) $\pi$ is positive on a neighborhood of $\psi_*$.
\end{assum}
For a correctly specified moment conditions model, the asymptotic normality of the BETEL posterior is established in the following theorem where we denote by $\pi(\sqrt{n}(\psi-\psi_*)|x_{1:n})$ the posterior distribution of $\sqrt{n}(\psi-\psi_*)$. The result shows that the BETEL posterior distribution has a Gaussian limiting distribution and that it concentrates on a $n^{-1/2}$-ball centered at the true value of the parameter. An informal discussion of this behavior is given by \cite{Schennach2005} but without the required assumptions. Theorem \ref{thm_BvM_correctly_specified} below provides these assumptions. The proof of the result is based on \textit{e.g.} \cite{LehmanCasella1998} and is given in the Supplementary Appendix.

\begin{thm}[Bernstein - von Mises -- correct specification]\label{thm_BvM_correctly_specified}
  Under Assumptions \ref{Ass_0_NS} - \ref{Ass_prior} and if in addition, for any $\delta > 0$, there exists an $\epsilon>0$ such that, as $n\rightarrow \infty$
  \begin{equation}\label{Ass_identification_rate_contraction_correct_specification}
    P\left(\sup_{\|\psi - \psi_*\| > \delta}\frac{1}{n}\sum_{i=1}^n \lt(l_{n,\psi}(x_i) - l_{n,\psi_*}(x_i)\rt) \leq - \epsilon\right) \rightarrow 1,
  \end{equation}
  \noindent then the posteriors converge in total variation towards a normal distribution, that is,
  \begin{equation}\label{Ass_identification_rate_contraction_correct_specification_2}
    \sup_{B}\left|\pi(\sqrt{n}(\psi-\psi_*)\in B|x_{1:n}) - \mathcal{N}_{0,\lt(\Gamma'\Delta^{-1}\Gamma\rt)^{-1}}(B)\right|\overset{p}{\to} 0
  \end{equation}
  \noindent where $B\subseteq \Psi$ is any Borel set.
\end{thm}
According to this result, the posterior distribution $\pi(\psi|x_{1:n})$ of $\psi$ is asymptotically normal, centered on the true value $\psi_*$ and with variance $n^{-1}\lt(\Gamma'\Delta^{-1}\Gamma\rt)^{-1}$. Thus, the posterior distribution has the same asymptotic variance as the efficient Generalized Method of Moments estimator of \cite{Hansen1982} (see also \cite{Chamberlain1987}). Assumption \eqref{Ass_identification_rate_contraction_correct_specification} in this theorem is a standard 
identifiability condition (see \textit{e.g.} \cite[Assumption 6.B.3]{LehmanCasella1998}) that controls the behavior of the log-ETEL function at a distance from $\psi_*$. Controlling this behavior is important because the posterior involves integration over the whole range of $\psi$. To understand the meaning of this assumption, we
remark that asymptotically the $\log$-ETEL function $\psi\mapsto\sum_{i=1}^n l_{n,\psi}(x_i)$ is maximized at the true value $\psi_*$ because the model is correctly specified. Hence, Assumption \eqref{Ass_identification_rate_contraction_correct_specification} means that if the parameter $\psi$ is ``far'' from the true value $\psi_*$ then the $\log$-ETEL function has to be small, that is, has to be far from the maximum value $\sum_{i=1}^n l_{n,\psi_*}(x_i)$.

%========================================================================================================================================
\subsection{Asymptotic Properties: misspecification}\label{s_Asymptotic_result_ms}
%\subsubsection{Asymptotic properties under misspecification}\label{sss_asymptotic_misspecification}
In this section, we consider the case where the model is misspecified in the sense of Definition \ref{def_misspecified_model} and establish that, even in this case, the BETEL posterior distribution has good frequentist asymptotic properties as the sample size $n$ increases. Namely, we show that the BETEL posterior of $\sqrt{n}(\psi - \psi_*)$ is asymptotically normal and the BETEL posterior of $\psi$ concentrates on a $n^{-1/2}$-ball centred at the pseudo-true value of the parameter. To the best of our knowledge, these properties have not been established yet for misspecified models.

Because in misspecified models there is no value of $\psi$ for which the true data distribution $P$ satisfies the restriction \eqref{eq_augmented_model}, we need to define a pseudo-true value for $\psi$. The latter is defined as the value of $\psi$ that minimizes the KL  divergence $K(P||Q^*(\psi))$ between the true data distribution $P$ and a distribution $Q^*(\psi)$ defined as $Q^*(\psi) := \arginf_{Q\in\mathcal{P}_{\psi}}K(Q||P)$, where $K(Q||P):=\int\log(dQ/dP)dQ$ and $\mathcal{P}_{\psi}$ is defined in Definition \ref{def_misspecified_model}. We remark that these two KL divergences are the population counterparts of the KL divergences used for the definition of the ETEL function in \eqref{eq_ETEL_function}: the empirical counterpart of $K(Q||P)$ is used to construct the $p_i^*(\psi)$ probabilities and is given by \eqref{eq:Entropy_0}, while the empirical counterpart of $K(P||Q^*(\psi))$ is given by $\log(1/n)-\sum_{i=1}^n l_{n,\psi}(x_i)/n$ where $\sum_{i=1}^n l_{n,\psi}(x_i)$ is the log-ETEL function if the dual theorem holds. Roughly speaking, the pseudo-true value is the value of $\psi$ for which the distribution that satisfies the corresponding restrictions \eqref{eq_augmented_model} is the closest to the true $P$, in the KL sense. By using the dual representation of the KL minimization problem, the $P$-density $dQ^*(\psi)/dP$ admits a closed-form: $dQ^*(\psi)/dP = e^{\lambda_\circ(\psi)'g^A(X,\psi)}/\EE^P\left[e^{ \lambda_\circ(\psi)'g^A(X,\psi)}\right]$ where $\lambda_\circ(\psi)$ is the pseudo-true value of the tilting parameter defined as the solution of $\EE^P[\exp\{\lambda'g^A(X,\psi)\}g^A(X,\psi)] = 0$ which is unique by the strict convexity of $\EE^P[\exp\{\lambda'g^A(X,\psi)\}]$ in $\lambda$. Therefore,
\begin{eqnarray}\label{eq_def_pseudo_true_value}
  \lambda_\circ(\psi) & := & \arg\min_{\lambda\in\mathbb{R}^d}\EE^P\left[e^{\lambda'g^A(X,\psi)}\right],\nonumber\\
 \psi_\circ & := & \arg\max_{\psi\in\Psi}\EE^P\log\left[\frac{e^{\lambda_\circ(\psi)'g^A(X,\psi)}}{\EE^P\left[e^{ \lambda_\circ(\psi)'g^A(X,\psi)}\right]}\right].\label{theta_misspecified}
\end{eqnarray}
\noindent However, in a misspecified model, the dual theorem is not guaranteed to hold and so $\psi_\circ$ defined in \eqref{eq_def_pseudo_true_value} is not necessarily equal to the pseudo-true value defined as the KL-minimizer. In fact, when the model is misspecified, the probability measures in $\mathcal{P}:=\bigcup_{\psi\in\Psi}\mathcal{P}_\psi$, which are implied by the model, might not have a common support with the true $P$, see \cite{Sueishi2013} for a discussion on this point. Following \cite[Theorem 3.1]{Sueishi2013}, in order to guarantee identification of the pseudo-true value by \eqref{theta_misspecified} and validity of the dual theorem we introduce the following assumption. This assumption replaces Assumption \ref{Ass_0_NS} in misspecified models.
\begin{assum}\label{Ass_absolute continuity}
  For a fixed $\psi\in\Psi$, there exists $Q\in \mathcal{P}_{\psi}$ such that $Q$ is mutually absolutely continuous with respect to $P$, where $\mathcal{P}_{\psi}$ is defined in Definition \ref{def_misspecified_model}.
\end{assum}
This assumption implies that $\mathcal{P}_{\psi}$ is non-empty. A similar assumption is also made by \citet{kleijn2012} to establish the BvM under misspecification. Moreover, because consistency in misspecified models is defined with respect to the pseudo-true value $\psi_\circ$, we need to replace Assumption \ref{Ass_prior} \textit{(b)} by the following assumption which, together with Assumption \ref{Ass_prior} \textit{(a)}, requires the prior to put enough mass to balls around $\psi_\circ$.
\begin{assum}\label{Ass_prior_BvM}
  The prior distribution $\pi$ is positive on a neighborhood of $\psi_\circ$ where $\psi_\circ$ is as defined in \eqref{eq_def_pseudo_true_value}.
\end{assum}
\indent In addition to these assumptions, to prove Theorem \ref{thm_BvM_misspecified} below we also use Assumptions \ref{Ass_1_NS} \textit{(a)}-\textit{(d)} and \ref{Ass_2_NS} \textit{(b)} in the previous section. Finally, in order to guarantee $n^{-1/2}$-convergence of $\wh\lambda$ towards $\lambda_\circ$ and $n^{-1/2}$-contraction of the posterior distribution of $\psi$ around $\psi_\circ$, we introduce Assumptions \ref{Ass_3_invalid_moments} and \ref{Ass_3_BvM}. These assumptions require the pseudo-true values $\lambda_\circ$ and $\psi_\circ$ to be in the interior of a compact parameter space, and the function $g^A(x,\psi)$ to be sufficiently smooth and uniformly bounded as a function of $\psi$. These assumptions are not new in the literature and are also required by \cite[Theorem 10]{Schennach2007} (adapted to account for the augmented model).
\begin{assum}\label{Ass_3_invalid_moments}
  (a) There exists a function $M(\cdot)$ such that $\EE^P[M(X)]<\infty$ and $\|g^A(x,\psi)\|\leq M(x)$ for all $\psi\in\Psi$;
  (b) $\lambda_\circ(\psi)\in int(\Lambda(\psi))$ where $\Lambda(\psi)$ is a compact set and $\lambda_{\circ}$ is as defined in \eqref{eq_def_pseudo_true_value}; (c) it holds $\EE^P\left[\sup_{\psi\in\Psi,\lambda\in\Lambda(\psi)}e^{\{\lambda^{\prime}g^A(X,\psi)\}}\right]<\infty.$
\end{assum}
%
%\noindent The next assumption guarantees $n^{-1/2}$-convergence of $\wh\lambda$ and of the ET-implied probabilities towards their pseudo-true value, see \cite[Theorem 10]{Schennach2007}, and is used to prove the BvM theorem under misspecification and the other results in section \ref{s_Asymptotic_result_ms} below. The result in the BvM, in turn, allows to prove Theorem \ref{thm_consistency_selection_misspecification}.
%
\begin{assum}\label{Ass_3_BvM}
  Let $\psi_{\circ}$ be as defined in \eqref{eq_def_pseudo_true_value}.
  (a) The pseudo-true value $\psi_\circ\in int(\Psi)$ is the unique maximizer of $$\lambda_\circ(\psi)'\EE^P[g^A(X,\psi)] - \log\EE^P[\exp\{\lambda_\circ(\psi)'g^A(X,\psi)\}],$$ where $\Psi$ is compact; (b) $S_{jl}(x_i,\psi) := \partial^2 g^A(x_i,\psi)/\partial \psi_j\partial \psi_l$ is continuous in $\psi$ for $\psi\in\mathcal{U}_\circ$, where $\mathcal{U}_\circ$ denotes a ball centred at $\psi_\circ$ with radius $n^{-1/2}$; (c) there exists $b(x_i)$ satisfying $\EE^P\left[\sup_{\psi\in\mathcal{U}_\circ}\sup_{\lambda\in\Lambda(\psi)}\exp\{\kappa_1\lambda'g^A(X,\psi)\}b(X)^{\kappa_2}\right]<\infty$ for $\kappa_1=0,1,2$ and $\kappa_2 = 0,1,2,3,4$ such that $\|g^A(x_i,\psi)\|<b(x_i)$, $\|\partial g^A(x_i,\psi)/\partial\psi'\|_F\leq b(x_i)$ and $\|S_{jl}(x_i,\psi)\|\leq b(x_i)$ for $j,l=1,\ldots,p$ for any $x_i\in (\mathcal{X},\mathfrak{B}_{\mathcal{X}})$ and for all $\psi\in\mathcal{U}_\circ$.
\end{assum}

\indent A first step to establish the BvM theorem is to prove that the misspecified model satisfies a stochastic Local Asymptotic Normality (LAN) expansion around the pseudo-true value $\psi_\circ$. Namely, that the log-likelihood ratio $l_{n,\psi} - l_{n,\psi_\circ}$, evaluated at a local parameter around the pseudo-true value, is well approximated by a quadratic form. Such a result is established in Theorem \ref{thm_stochastic_LAN} in the Appendix. A second key ingredient for establishing the BvM theorem is the requirement that, as $n\rightarrow\infty$, the posterior of $\psi$ concentrates and puts all its mass on $\Psi_n := \left\{\|\psi - \psi_\circ\|\leq M_n/\sqrt{n}\right\}$, where $M_n$ is any sequence such that $M_n\rightarrow\infty$. We prove this result in Theorem \ref{thm_posterior_consistency_misspecified} in the Appendix.\\
\indent Theorem \ref{thm_BvM_misspecified} states that the limit of the posterior distribution of $\sqrt{n}(\psi - \psi_\circ)$ is a Gaussian distribution with mean and variance defined in terms of the population counterpart of $l_{n,\psi}(x)$, which we denote by $\mathfrak{L}_{n,\psi}(x):=\log\frac{\exp(\lambda_\circ(\psi)'g^A(x,\psi))}{\EE^P[\exp(\lambda_\circ(\psi)'g^A(x,\psi))]} - \log n$ and which involves the pseudo-true value $\lambda_\circ$. With this notation, the variance and mean of the Gaussian limiting distribution are $V_{\psi_\circ}^{-1} := - (\EE^P[\ddot{\mathfrak{L}}_{n,\psi_\circ}])^{-1}$ and $\Delta_{n,\psi_\circ} := \frac{1}{\sqrt{n}}\sum_{i=1}^nV_{\psi_\circ}^{-1}\dot{\mathfrak{L}}_{n,\psi_\circ}(x_i)$, respectively, where $\dot{\mathfrak{L}}_{n,\psi_\circ}$ and $\ddot{\mathfrak{L}}_{n,\psi_\circ}$ denote the first and second derivatives of the function $\psi\mapsto \mathfrak{L}_{n,\psi}$ evaluated at $\psi_\circ$. Let $\pi(\sqrt{n}(\psi-\psi_\circ)|x_{1:n})$ denote the posterior distribution of $\sqrt{n}(\psi-\psi_\circ)$.

\begin{thm}[Bernstein - von Mises -- misspecification]\label{thm_BvM_misspecified}
  Assume that the matrix $V_{\psi_\circ}$ is nonsingular and that Assumptions \ref{Ass_1_NS} \textit{(a)}-\textit{(d)}, \ref{Ass_2_NS} \textit{(b)}, \ref{Ass_prior} \textit{(a)}, \ref{Ass_absolute continuity}, \ref{Ass_prior_BvM}, \ref{Ass_3_invalid_moments}, and \ref{Ass_3_BvM} hold. If in addition there exists a constant $C>0$ such that for any sequence $M_n\rightarrow \infty$, as $n\rightarrow \infty$
  \begin{equation}\label{Ass_identification_rate_contraction}
    P\left(\sup_{\psi\in \Psi_n^c}\frac{1}{n}\sum_{i=1}^n \lt(l_{n,\psi}(x_i) - l_{n,\psi_\circ}(x_i)\rt) \leq - \frac{CM_n^2}{n}\right) \rightarrow 1,
  \end{equation}
  \noindent then the posteriors converge in total variation towards a normal distribution, that is,
  \begin{equation}\label{eq_BvM_missspecification}
    \sup_{B}\left|\pi(\sqrt{n}(\psi-\psi_\circ)\in B|x_{1:n}) - \mathcal{N}_{\Delta_{n,\psi_\circ},V_{\psi_\circ}^{-1}}(B)\right|\overset{p}{\to} 0
  \end{equation}
  \noindent where $B\subseteq \Psi$ is any Borel set.
\end{thm}
Condition \eqref{Ass_identification_rate_contraction} involves the log-likelihood ratio $l_{n,\psi}(x) - l_{n,\psi_\circ}(x)$ and is an identifiability condition, standard in the literature, and with a similar interpretation as condition \eqref{Ass_identification_rate_contraction_correct_specification}. Theorem \ref{thm_BvM_misspecified} states that, in misspecified models, the sequence of posterior distributions converges in total variation to a sequence of normal distributions with random mean and fixed covariance matrix $V_{\psi_\circ}^{-1}$. By using the first order condition for $\psi_\circ$ it can be shown that the random mean $\Delta_{n,\psi_\circ}$ has mean zero. %Unlike Theorem \ref{thm_BvM_correctly_specified} for correctly specified models, in Theorem \ref{thm_BvM_misspecified} the centering $\Delta_{n,\psi_\circ}$ of the limiting distribution is in general non-zero since $\lambda_\circ\neq 0$.
We stress that the BvM result of Theorem \ref{thm_BvM_misspecified} for the BETEL posterior distribution does not directly follow from the assumptions and results in \cite{kleijn2012} because the ETEL function contains random quantities. Therefore, we need to strengthen the assumptions in order to establish that a stochastic LAN expansion holds for our case.

As the next lemma shows, the quantity $\Delta_{n,\psi_\circ}$ relates to the \cite{Schennach2007}'s ETEL frequentist estimator $\wh\psi$ (whose definition is recalled in \eqref{eq_ETEL_def} in the Appendix for convenience). Because of this connection, it is possible to write the location of the normal limit distribution in a more familiar form in terms of the semi-parametric efficient frequentist estimator $\wh\psi$.
\begin{lem}\label{lem_asymptotic_expansion_ETEL_misspecified}
  Assume that the matrix $V_{\psi_\circ}$ is nonsingular and that Assumptions \ref{Ass_1_NS} \textit{(a)}-\textit{(d)}, \ref{Ass_2_NS} \textit{(b)}, \ref{Ass_absolute continuity}, \ref{Ass_3_invalid_moments}, and \ref{Ass_3_BvM} hold. Then, the ETEL estimator $\wh\psi$ satisfies
  \begin{equation}
    \sqrt{n}(\wh\psi - \psi_\circ) = \frac{1}{\sqrt{n}}\sum_{i=1}^n V_{\psi_\circ}^{-1}\dot{\mathfrak{L}}_{n,\psi_\circ} + o_p(1).
  \end{equation}
\end{lem}
Therefore, Lemma \ref{lem_asymptotic_expansion_ETEL_misspecified} implies that the BvM Theorem \ref{thm_BvM_misspecified} can be reformulated with the sequence $\sqrt{n}(\wh \psi - \psi_\circ)$ as the location of the normal limit distribution, that is,
\begin{equation}\label{eq_BvM_alternative}
  \sup_{B}\left|\pi(\psi\in B |x_{1:n}) - \mathcal{N}_{\wh\psi,n^{-1}V_{\psi_\circ}^{-1}}(B)\right| \overset{p}{\to}0.
\end{equation}
Two remarks are in order: (I) the limit distribution of $\sqrt{n}(\wh\psi - \psi_\circ)$ is centred on zero because $\EE^P[\dot{\mathfrak{L}}_{n,\psi_\circ}]= 0$% at the rate $n^{-1/2}$
; (II) the asymptotic covariance matrix of $\sqrt{n}(\wh\psi - \psi_\circ)$ is $V_{\psi_\circ}^{-1}\EE^P[\dot{\mathfrak{L}}_{n,\psi_\circ}\dot{\mathfrak{L}}_{n,\psi_\circ}']V_{\psi_\circ}^{-1}$ (which is also derived in \cite[Theorem 10]{Schennach2007}) and, because of misspecification, it does not coincide with the limiting covariance matrix in the BvM theorem. This consequence of misspecification is also discussed in \cite{kleijn2012} and implies that, for $\alpha\in (0,1)$, the central $(1 - \alpha)$ Bayesian credible sets are not in general $(1-\alpha)$ confidence sets, even asymptotically. In fact, while credible sets are correctly centered, their width/volume need not be correct since the asymptotic variance matrix in the BvM is not the sandwich asymptotic covariance matrix of the frequentist estimator.

\begin{example}[Misspecified model and pseudo-true value]
  Let us consider the model $y_i = \alpha + e_i$, $i=1,\ldots,n$, with $e_i$ independently drawn from the skewed distribution $P$ given in \eqref{eq:dgp3}. We consider the following two moment conditions $\EE^{P}[y_{i}-\alpha] = 0$ and $\EE^{P}[(y_{i}-\alpha)^{3}] =0$. This situation is different from the one illustrated in Example 1 because there are no covariates and the augmented parameter $v$ is (incorrectly) forced to be zero. In turn, $\alpha$ has to satisfy both the moment restrictions, which is impossible under $P$. Instead, for each $\alpha$ the ETEL likelihood function is defined by the probability measure $Q^{*}(\alpha)$ which is the closest to the true generating process $P$ in terms of KL divergence among the probability measures that are consistent with the given moment restrictions for a given $\alpha$. In Figure \ref{fig:betelpost2} (left panel), we present $\EE^{P}[\log (dQ^*(\alpha)/dP)]$ which is equal to $-K(P||Q^*(\alpha))$. The value that maximizes this function is different from the true value $(\alpha=0)$ and it is peaked around $-0.056$. This value is the pseudo-true value. In the right panel of Figure \ref{fig:betelpost2}, we present the BETEL posterior distribution of $\alpha$ for five different sample sizes. The BETEL posterior distribution shrinks and moves toward the pseudo-true value, in conformity with our theoretical result.
\begin{figure}[h!]
     \centering
\begin{subfigure}[b]{1\textwidth}
        \resizebox{\linewidth}{!}{
           \input{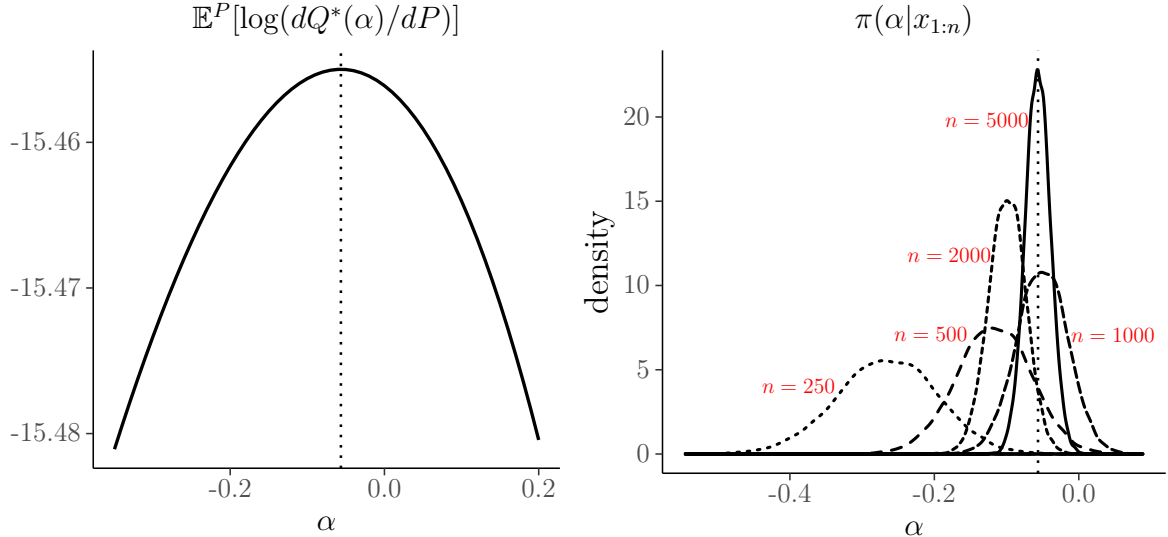}
        }
    \end{subfigure}
    \caption{\small{Posterior distributions in Example 2 under misspecification. Left panel presents the function $\alpha \mapsto\EE^{P}[\log (dQ^*(\alpha)/dP)]$ where $Q^*(\alpha) := \arginf_{Q\in\mathcal{P}_{\psi}}K(Q||P)$ with $\psi := (\alpha,0)$. For each $\alpha$, we approximate this function based on the dual representation in \eqref{theta_misspecified} -- which is valid under Assumption \ref{Ass_absolute continuity} -- using five million simulation draws from $P$. In the right panel, we present the BETEL posterior distribution of the location parameter $\alpha$ for $n=250, 500, 1000, 2000, 5000$ where $n$ is the number of observations. The prior distribution for $\alpha$ is student-t with mean 0 and dispersion 5. Vertical dashed lines are at the pseudo-true parameter value, approximately equal to $-0.056$.  Posterior densities are based on 10,000 draws beyond a burn-in of 1000. The M-H acceptance rate is about 90\% for each sample size.}}
    \label{fig:betelpost2}
\end{figure}
\end{example}

%==========================================================================
\section{Bayesian Model Selection}\label{s_BMS}

\subsection{Basic idea}\label{ss_Basic_idea}
Now suppose that there are countable candidate models indexed by $\ell$. Suppose that model $\ell$ is characterized by
\bee
\EE^{P}[g^{\ell}(X,\theta^{\ell})] = 0,
\eee
with $\theta^{\ell} \in \Theta^{\ell} \subset \mathbb{R}^{p_{\ell}}$, and $\ell = 1,\ldots, J$ for some $J\geq 2$. Different models involve different parameters of interest $\theta^\ell$ and/or different $g^\ell$ functions. If $X$ contains a dependent variable and covariates it might be that the covariates are not the same for all models, however to lighten the notation we do not explicit this difference across the models.

One or all models may be misspecified. The goal is to compare these models and select the best model. By best model we mean the model that contains the maximum number of over-identifying conditions when all models are correctly specified, and when all models are misspecified, we mean the model that is the closest to the true $P$. Our purpose in this section is to establish a collection of results on the search for such a best model. We show that this search can be carried out with the help of the marginal likelihoods (defined as the integral of the sampling density over the parameters with respected to the prior density) of the competing models. The model with the largest marginal likelihood satisfies a model selection consistency property in that the model chosen in this way is the best model asymptotically. This property, which has not been established in this context before, is of enormous practical and theoretical importance.

Before getting to the details, it is crucial to understand that there are some subtleties involved in comparing different moment condition models. The central problem is that the marginal likelihood of models with different sets of moment restrictions and different parameters may not be comparable. In fact, when we have different sets of moment restrictions, we need to be careful about dealing with, and interpreting, unused moment restrictions. This can be best explained by an example.

\setcounter{example}{0}
\begin{example}[continued]
  Suppose we do not know if $e_{i}$ is symmetric. In this case, one might be inclined to compare the following two candidate models:
\bee \label{eq:modelprime}
\begin{split}
\text{Model 1} &:  \EE^{P}[e_{i}(\theta) ] = 0, \quad \EE^{P}[e_{i}(\theta)z_{i}]=0.\\
\text{Model 2} &:  \EE^{P}[e_{i}(\theta) ] = 0, \quad \EE^{P}[e_{i}(\theta)z_{i}]=0, \quad \EE^{P}[(e_{i}(\theta))^{3}] =0.
\end{split}
\eee
where $\theta = (\alpha, \beta)$ is the same parameter in the two models and $e_{i}(\theta) = (y_{i} - \alpha - \beta z_{i})$. As written, these two models are not comparable because the convex hulls associated with the two models do not have the same dimension. More precisely, let $co_1:=\{\sum_{i=1}^n p_i (e_i(\theta),e_i(\theta)z_i)'; p_i\geq 0, \forall i=1,\ldots,n, \, \sum_{i=1}^n p_i = 1\}$ be the convex hull associated with Model $1$, and $co_2:=\{\sum_{i=1}^n p_i (e_i(\theta),e_i(\theta)z_i,e_i(\theta)^3)'; p_i\geq 0, \forall i=1,\ldots,n, \, \sum_{i=1}^n p_i = 1\}$ be the convex hull associated with Model $2$. Because $co_1$ and $co_2$ have different dimensions, the $p_i^*(\theta)$ in the two ETEL functions are not comparable because they enforce the zero vector constraint (the second constraint in \eqref{eq: Entropy}) in different spaces ($\mathbb{R}^2$ and $\mathbb{R}^3$).
\end{example}

The foregoing problem can be overcome as follows. We start by defining a grand model that nests all the models that we want to compare. This grand model is constructed such that: (1) it includes all the moment restrictions in the models and, (2) if the same moment restriction is included in two or more models but involves a different parameter in different models, then the grand model includes the moment restriction that involves the parameter of largest dimension. We write the grand model as $\EE^{P}[g^{G}(X, \theta^{G})]=0$ where $g^G$ has dimension $d$, and $\theta^G$ includes the parameters of all models. Next, each original model is obtained from this grand model by first subtracting a vector of nuisance parameters $V$ and then restricting $\theta^G$ and $V$ appropriately. More precisely, an equivalent version of the original model is obtained by: (I) setting equal to zero the components of $\theta^G$ in the shared moment restrictions that are not present in the original model, (II) letting free the components of $V$ that correspond to the over-identifying moment restrictions not present in the original model and, (III) setting equal to zero the components of $V$ that correspond to moment restrictions present in the original model and to moment restrictions that exactly identify the extra parameters arising from the other models.

The set of models that emerge from this strategy are equivalent to the original collection but, crucially, are now comparable. Also importantly for practice, as our illustrations of this strategy show below, the strategy just described is simple to operationalize. With this formulation, model $\ell$, denoted by $M_\ell$, is then defined from the grand model as
\bee\label{eq_augmented_grand_model}
  \EE^{P}[g^{A}(X, \theta^{\ell}, v^{\ell})] = 0,\qquad \theta^{\ell} \in \Theta^{\ell} \subset \mathbb{R}^{p_{\ell}}
\eee
\noindent where $g^A(X, \theta^{\ell}, v^{\ell}) = g^G(X, \theta^{\ell}) - V^{\ell}$ with $V^\ell\in \mathfrak{V}\subset\mathbb{R}^d$ and with $v^{\ell} \in \mathcal{V}^{\ell} \subset \mathbb{R}^{d_{v_{\ell}}}$ being the vector that collects all the non-zero components of $V^\ell$. We assume that $0\leq d_{v_{\ell}} \leq d - p_{\ell}$ in order to guarantee identification of $\theta^\ell$. The parameter $v^\ell$ is the augmented parameter and $\theta^\ell$ is the parameter of interest for model $\ell$ that has been obtained from $\theta^G$ by doing the transformation in (I). Hereafter, we use the notation $\psi^\ell := (\theta^\ell,v^\ell)\in \Psi^\ell$ with $\Psi^\ell:=\Theta\times\mathcal{V}^\ell$.

\setcounter{example}{0}
\begin{example}[continued]
  To be able to compare Model 1 and Model 2 in \eqref{eq:modelprime}, we construct the grand model as $\EE^P[g^G(x_i,\theta)]:=\EE^P[(e_{i}(\theta),e_{i}(\theta)z_{i},e_{i}(\theta)^3)']$. With respect to this grand model, Model 1 and Model 2 are reformulated as $M_1$ and $M_2$, respectively, by applying (II) and (III) above:
\bee
\begin{split}
M_{1}&:  \EE^{P}[e_{i}(\theta) ] = 0, \quad \EE^{P}[e_{i}(\theta)  z_{i}]=0, \quad \EE^{P}[(e_{i}(\theta))^{3}] = v\\
M_{2}&:  \EE^{P}[e_{i}(\theta) ] = 0, \quad \EE^{P}[e_{i}(\theta)z_{i}]=0, \quad \EE^{P}[(e_{i}(\theta))^{3}] =0.
\end{split}
\eee
So, $\psi^1 = (\theta',v)'$ and $\psi^2 = \theta$. The convex hulls of $M_1$ and $M_2$ have both dimension $3$ and more importantly, if $co(M_1)$ and $co(M_2)$ denote these two convex hulls, we have $co(M_1) = co(M_2) - V$ where $V = (0,0,v)'$ so that the two models are comparable. It is important to note how Model 1 in \eqref{eq:modelprime} and $M_{1}$ deal with uncertainty about the third moment restriction: Model 1 in \eqref{eq:modelprime} ignores its uncertainty completely while $M_{1}$ models the degree of uncertainty through the augmented parameter $v$. This argument is not limited to comparing two models. When we have multiple models, we need to make sure that the grand model encompasses all candidate models through augmented parameters.
\end{example}

We note that this strategy covers both nested and non-nested models. We say that two models are non-nested, in their original formulation, if neither model can be obtained from the other by eliminating some moment restriction, or by setting to zero some parameter, or both. Points (II) and (III) above are important for the treatment of such non-nested models. In fact, points (II) and (III) imply that, if there are moment restrictions not present in the original model that involve parameters that are not in the original model, then a number of these extra moment restrictions equal to the number of the extra parameters has to be included. This does not alter the original model if these extra moment restrictions exactly identify the extra parameters and so place no restrictions on the data generating process. Moreover, despite the notation, for non-nested models $\theta^\ell$ in \eqref{eq_augmented_grand_model} might be larger than the parameter in the original model $\ell$.

In what follows, we show how to compute the marginal likelihood for a model. Then, in Section \ref{ss_consistency_ML} we formally show that, with probability approaching one as the number of observations
increases, the marginal likelihood based selection procedure favors the model with the minimum number of parameters of interest and the maximum number of valid moment restrictions. We also consider the situation where all models are misspecified. In this case, our model selection procedure selects the model that is closer to the true data generating process in terms of the KL divergence.

\subsection{Marginal Likelihood}\label{ss_ML}

For each model $M_{\ell }$, we impose a prior distribution for $\psi^\ell$ on $\Psi^{\ell }$, and obtain the BETEL posterior distribution based on \eqref{eq:BETELpost}. Then, we select the model with the largest marginal likelihood, denoted by $m(x_{1:n};M_{\ell })$, which we calculate by the method of \cite{Chib1995} as extended to Metropolis-Hastings samplers in \cite{ChibJeliazkov2001}. This method makes computation of the marginal likelihood simple and is a key feature of our procedure. The main advantage of the \cite{Chib1995} method is that it is calculable from the same inputs and outputs that are used in the MCMC sampling of the posterior distribution. The starting point of this method is the following identity of the log-marginal likelihood introduced in \cite{Chib1995}
\begin{equation}
\log m(x_{1:n};M_{\ell })=\log \pi (\tilde\psi^\ell|M_{\ell})+\log p(x_{1:n}|\tilde\psi^\ell,M_{\ell })-\log \pi (\tilde\psi^\ell|x_{1:n},M_{\ell }),  \label{eq_MSC_ML}
\end{equation}%
where $\tilde\psi^\ell$ is any point in the support of the posterior (such as the posterior mean) and the dependence on the model $M_{\ell }$ has been made explicit. The first two terms on the right-hand side of this decomposition are available directly whereas the third term can be estimated from the output of the MCMC simulation of the BETEL posterior distribution. For example, in the context of the one block MCMC algorithm given in Section \ref{ss_prior_posterior}, from \cite{ChibJeliazkov2001}, we have that
\begin{equation*}
\pi (\tilde\psi^\ell|x_{1:n},M_{\ell })=\frac{\EE_{1}\left\{\alpha \left( {\psi^\ell},\tilde\psi^\ell|x_{1:n},M_{\ell}\right) q(\tilde\psi^\ell)x_{1:n},M_{\ell })\right\} }{\EE_{2}\left\{ \alpha (\tilde\psi^\ell,\psi^\ell|x_{1:n},M_{\ell })\right\} }
\end{equation*}%
where $\EE_{1}$ is the expectation with respect to $\pi (\psi^\ell|x_{1:n},M_{\ell })$ and $\EE_{2}$ is the expectation with respect to $q(\psi^\ell|x_{1:n},M_{\ell })$. These expectations can be easily approximated by simulations.

%===========================================================================================================================================
\subsection{Model selection consistency results}\label{ss_consistency_ML}

In this section we establish the consistency of our marginal likelihood based selection procedure for the following exhaustive cases: the case where the models in the comparison set contain only valid moment restrictions, the case where all the models in the set are misspecified, and finally the case where some of the models contain only valid moment restrictions while the others contain at least one invalid moment restriction. Our proofs of consistency are based on: (I) the results of the BvM theorems for correctly and misspecified models stated in Sections \ref{s_Asymptotic_result_cs} and \ref{s_Asymptotic_result_ms}, and (II) the analysis of the asymptotic behavior of the ETEL function under correct and misspecification which we develop in the Appendix (see Lemmas \ref{lem_likelihood_ET_approx} and \ref{Lemma_Uniform_conv_misspecified}).

The first theorem states that, if the active moment restrictions are all valid, then the marginal likelihood selects the model that contains the maximum number of overidentifying conditions, that is, the model with the maximum number of active moment restrictions and the smallest number of parameters of interest. This means that the marginal likelihood based selection procedure enforces parsimony.

For a model $M_\ell$, the dimension of the parameter of interest $\theta^{\ell}$ to be estimated is $p_\ell$ while the number of active moment restrictions (included in the model for the estimation of $\theta^\ell$) is $(d-d_{v_\ell})$. Consider two generic models $M_1$ and $M_2$. Then, $d_{v_2} < d_{v_1}$ means that model $M_{2}$ contains more active restrictions than model $M_{1}$, and $p_{2}<p_{1}$ means that model $M_{1}$ contains more parameters of interest to be estimated than $M_{2}$.

\begin{thm}\label{thm_consistency_correct_specification_2}
  Let Assumptions \ref{Ass_1_NS} -- \ref{Ass_prior} and \eqref{Ass_identification_rate_contraction_correct_specification} hold, and consider $J<\infty$ different models $M_{\ell}$, $\ell = 1,\ldots,J$, that satisfy Assumption \ref{Ass_0_NS}, that is, they are all correctly specified. Then,
  \begin{displaymath}
    \lim_{n\rightarrow \infty}P\left(\max_{\ell\neq j}\log m(x_{1:n};M_{\ell}) < \log m(x_{1:n};M_{j})\right) = 1
  \end{displaymath}
  if and only if $p_{j} + d_{v_j} < p_{\ell} + d_{v_{\ell}}$, $\forall \ell\neq j$.
%  \noindent This implies that $B_{12}<1$ with probability approaching $1$.
\end{thm}
\noindent The result of the theorem implies that, with probability approaching $1$, the Bayes factor $B_{j\ell}: = m(x_{1:n};M_{j})/m(x_{1:n};M_{\ell})$ is larger than $1$ for every $\ell \neq j$. The result in the theorem is an equivalence result saying that, if we compare models that contain only valid moment restrictions, then the marginal likelihood selects a model $M_{j}$ if and only if $M_j$ contains the maximum number of overidentifying conditions among all the compared models.

\setcounter{example}{2}
\begin{example}[Model selection when all models are correctly specified]\label{Example_4}
  We suppose that for every $i = 1,\ldots,n$, $y_i=\alpha + \beta z_i + e_i$, where $z_{i} \sim \mathcal{N}(0.5,1)$ and $e_{i}\sim \mathcal{N}(0,1)$ independently of $z_i$. Let $\theta = (\alpha,\beta)$, $e_{i}(\theta) = (y_{i} - \alpha - \beta z_{i})$ and the true value of $\theta$ be $(0,1)$. We compare the following models. Model 1: $\EE^P[(e_{i}(\theta),e_{i}(\theta)z_i,e_{i}(\theta)^3,e_{i}(\theta)^2 - 1)'] = 0$, Model 2: $\EE^P[(e_{i}(\theta),e_{i}(\theta)z_i,e_{i}(\theta)^2 - 1)'] = 0$ and Model 3: $\EE^P[(e_{i}(\theta),e_{i}(\theta)^2 - 1)']=0$ which, reformulated in terms of an encompassing grand model, become respectively:
\bee \label{eq:ex_moments}
\begin{split}
M_{1}&:  \EE^{P}[e_{i}(\theta) ] = 0, \quad \EE^{P}[e_{i}(\theta)z_{i}]=0, \quad \EE^{P}[e_{i}(\theta)^{3}] =0, \quad \EE^{P}[e_{i}(\theta)^{2} - 1] = 0\\
M_{2}&:  \EE^{P}[e_{i}(\theta) ] = 0, \quad \EE^{P}[e_{i}(\theta)z_{i}]=0, \quad \EE^{P}[e_{i}(\theta)^{3}] =v_{1}, \quad \EE^{P}[e_{i}(\theta)^{2} - 1] = 0,\\
M_{3}&:  \EE^{P}[e_{i}(\theta) ] = 0, \quad \EE^{P}[e_{i}(\theta)z_{i}]= v_{2}, \quad \EE^{P}[e_{i}(\theta)^{3}] = v_{1}, \quad \EE^{P}[e_{i}(\theta)^{2} - 1] = 0. \\
\end{split}
\eee
with $\psi^1 = \theta$, $\psi^2 = (\theta,v_1)$ and $\psi^3 = (\theta,v_1,v_2)$. Note that the last two moment restrictions (which concern the third and second moments) serve as extra information to infer the parameter $\theta$, when they are active. Under the standard normal error distribution, all the three models are correctly specified: $M_{1}$ has four active moment restrictions while $M_{2}$ and $M_{3}$ have three and two active moment restrictions, respectively.\\
\indent In Table \ref{tab:location_case1}, we report the percentage of times the marginal likelihood selects each of these models in $500$ trials, for different sample sizes. Model $M_1$, the model with the larger number of valid restrictions, is selected $99\%$ of times by sample size of $n=500$. The results are virtually indistinguishable for the training sample prior (based on 50 prior samples). Under both priors the proportion of correct selection tends to one.
\begin{table}[h!]
  \centering
  \begin{tabular}{lccclccc}
    \toprule
     & \multicolumn{3}{c}{\underline{Default prior}} & & \multicolumn{3}{c}{\underline{Training sample prior}} \\
     Model & $M_{1}$ & $M_{2}$ & $M_{3}$ & & $M_{1}$ & $M_{2}$ & $M_{3}$ \\
    \midrule
    $n=250$ & 97.8    & 1.6     & 0.6       &         & 98.0     & 1.6     & 0.4 \\
    $n=500$ & 99.0    & 0.8     & 0.2       &         & 99.0     & 0.8     & 0.2 \\
    $n=1000$ & 99.2    & 0.6     & 0.2       &         & 99.2    & 0.6      & 0.0 \\
    $n=2000$ & 99.2    & 0.8     & 0.0       &         & 99.2    & 0.8     & 0.0 \\
    \bottomrule
    \end{tabular}%
 \caption{\small{Model selection when all models are correctly specified. Frequency (\%) of times each of the three models in Example \ref{Example_4} are selected by the marginal likelihood criterion in $500$ trials, by sample size, for two different prior distributions.}}
\label{tab:location_case1}
\end{table}%
\end{example}

Next, we consider the case where all models are wrong in the sense of Definition \ref{def_misspecified_model}
and establish a major result of enormous practical significance. The result states that if we compare $J$ misspecified models, then the marginal likelihood based selection procedure selects the model with the smallest KL divergence $K(P||Q^*(\psi^{\ell}))$ between $P$ and $Q^*(\psi^\ell)$, where $Q^*(\psi^\ell)$ is such that $K(Q^*(\psi^\ell)||P) = \inf_{Q\in\mathcal{P}_{\psi^{\ell}}}K(Q||P)$ and $dQ^*(\psi^\ell)/dP = e^{\lambda_\circ(\psi)'g^A(X,\psi)}/\EE^P\left[e^{\lambda_\circ(\psi)'g^A(X,\psi)}\right]$ %with the second equality holding
by the dual theorem, as defined in Section \ref{s_Asymptotic_result_ms}. Because the I-projection $Q^*(\psi^\ell)$ on $\mathcal{P}_{\psi^{\ell}}$ is unique (\cite{csiszar1975}), which $Q^*(\psi^\ell)$ is closer to $P$ (in terms of $K(P||Q^*(\psi^{\ell}))$) depends only on the ``amount of misspecification'' contained in each model $\mathcal{P}_{\psi^{\ell}}$.

\begin{thm}\label{thm_consistency_selection_misspecification_2}
  Let Assumptions \ref{Ass_1_NS} - \ref{Ass_3_BvM} and \eqref{Ass_identification_rate_contraction} be satisfied. Let us consider the comparison of $J<\infty$ models $M_j$, $j=1,\ldots,J$ that all use misspecified moments, that is, $M_{j}$ does not satisfy Assumption \ref{Ass_0_NS}, $\forall j$. Then,
  \begin{displaymath}
    \lim_{n\rightarrow \infty}P\left(\log m(x_{1:n};M_{j}) > \max_{\ell\neq j}\log m(x_{1:n};M_{\ell})\right) = 1
  \end{displaymath}
  if and only if $K(P||Q^*(\psi^j))< \min_{\ell\neq j}K(P||Q^*(\psi^{\ell}))$, where $K(P||Q):=\int \log(dP/dQ)dP$.
\end{thm}
Similar as in Theorem \ref{thm_consistency_correct_specification_2}, Theorem \ref{thm_consistency_selection_misspecification_2} establishes the equivalence result that, if we compare models that all use misspecified moments, then the marginal likelihood selects a model $M_j$ if and only if $M_j$ has the smallest Kullback-Leibler divergence $K(P||Q^*(\psi^{j}))$ between the true data distribution $P$ and $Q^*(\psi^{j})$.
Remark that the condition $K(P||Q^*(\psi^j))<K(P||Q^*(\psi^{\ell}))$, $\forall \ell \neq j$, given in the theorem does not depend on a particular value of $\psi^j$ and $\psi^{\ell}$. Indeed, the result of the theorem hinges on the fact that the marginal likelihood selects the model with the $Q^*(\psi^j)$ the closer to $P$, that is, the model that contains the ``less misspecified'' moment restrictions for every value of $\psi^j$.

The result of the theorem also applies to the case where we compare a correctly specified model $M_1$ to misspecified models. Indeed, if model $M_1$ is correctly specified then $K(P||Q^*(\psi^1)) = 0$ while if model $M_j$ is misspecified then $K(P||Q^*(\psi^j)) > 0$.

\setcounter{example}{3}
\begin{example}[Model selection when all models are misspecified]\label{Example_misspecified_models}
  Recall that for $i=1,\ldots,n$, $y_i = \alpha + \beta z_i + e_i$. Here, we generate $z_{i} \sim \mathcal{N}(0.5,1)$ and $e_{i}$ from the skewed distribution in \eqref{eq:dgp3} with mean zero and variance $1.625$, independently of $z_i$. Let $\theta = (\alpha,\beta)'$, $e_{i}(\theta) = (y_{i} - \alpha - \beta z_{i})$ and the true value of $\theta$ be $(0,1)'$. We compare the following models. Model 4: $\EE^P[(e_{i}(\theta),e_{i}(\theta)z_i,e_{i}(\theta)^3,e_{i}(\theta)^2 - 2)'] = 0$, Model 5: $\EE^P[(e_{i}(\theta),e_{i}(\theta)z_i,e_{i}(\theta)^2 - 2)'] = 0$ and Model 6: $\EE^P[e_{i}(\theta),e_{i}(\theta)^2 - 2] = 0$ which, written in terms of an encompassing grand model, become respectively:
    \bee
    \begin{split}
    M_{4}&: \EE^{P}[e_{i}(\theta) ] = 0, \quad \EE^{P}[e_{i}(\theta)z_{i}]=0, \quad \EE^{P}[(e_{i}(\theta))^{3}] =0, \quad \EE^{P}[(e_{i}(\theta))^{2}-2] = 0 \\
    M_{5}&: \EE^{P}[e_{i}(\theta) ] = 0, \quad \EE^{P}[e_{i}(\theta)z_{i}]=0, \quad \EE^{P}[(e_{i}(\theta))^{3}] =v_{1}, \quad \EE^{P}[(e_{i}(\theta))^{2}-2] = 0 \\
    M_{6}&: \EE^{P}[e_{i}(\theta) ] = 0, \quad \EE^{P}[e_{i}(\theta)z_{i}]= v_{2}, \quad \EE^{P}[(e_{i}(\theta))^{3}] =v_{1}, \quad \EE^{P}[(e_{i}(\theta))^{2}-2] = 0
    \end{split}
    \eee
  \noindent with $\psi^4 = \theta$, $\psi^5 = (\theta,v_1)'$ and $\psi^6 = (\theta,v_1,v_2)'$. Thus, compared to Example \ref{Example_4}, here we change the moment restriction that involves the variance of $e_i$. When the underlying distribution has variance different from 2, all models $M_{4}$, $M_{5}$, and $M_{6}$ are misspecified due to the new moment restriction: $\EE^{P}[(e_{i}(\theta))^{2}-2] = 0$. In Table \ref{tab:location_case3}, we report the percentage of times the marginal likelihood selects each model out of 500 trials, by sample size, under the default and training sample prior (based on 50 prior observations).

  Because we know the true data generating process, we can compute, for each model, the KL divergence between the true model $P$ and $Q^{*}(\psi_{\circ}^j)$ at the pseudo-true parameter $\psi_{\circ}^j$ for model $M_j$ based on \eqref{theta_misspecified}. Using $10,000,000$ simulated draws from $P$, our calculations show that $K(P||Q^{*}(\psi_{\circ}^j))$ is equal to $0.0283$ for $M_{4}$, $0.0096$ for $M_{5}$, and $1.4901\times10^{-13}$ for $M_{6}$. Intuitively, $M_6$ is the closest to the true model since it imposes fewer restrictions (only two moment restrictions are active). This means that the set of probability distributions that satisfy $M_6$ is larger than (and contains) the sets of probabilities that conform with $M_4$ and $M_5$. This flexibility ensures that the divergence between the set of probabilities that satisfy $M_6$ and $P$ (as measured by the KL) will be at least as small as for $M_4$ and $M_5$. As the empirical results show, under each prior, the best model $M_6$ picked out by our marginal likelihood ranking is also the model that is the closest to the true model, consistent with the prediction of our theory.
  \begin{table}[h!]
  \centering
  \begin{tabular}{lccclccc}
    \toprule
     & \multicolumn{3}{c}{\underline{Default prior}} & & \multicolumn{3}{c}{\underline{Training sample prior}} \\
     Model & $M_{1}$ & $M_{2}$ & $M_{3}$ & & $M_{1}$ & $M_{2}$ & $M_{3}$ \\
    \midrule
    $n=250$ & 2.6    & 56.8    & 40.5 & & 3.0 & 62.0 & 35.0 \\
    $n=500$ & 0.2    & 28.1    & 71.6 & & 1.0 & 31.0 & 68.0 \\
    $n=1000$ & 0.0     & 4.2    & 95.8 & & 0.0 & 4.0 & 96.0 \\
    $n=2000$ & 0.0     & 0.0    & 100.0 & & 0.0 & 0.0 & 100.0 \\
    \bottomrule
    \end{tabular}
    \caption{\small{Model selection when all models are misspecified. Frequency (\%) of times each of the three models in Example \ref{Example_misspecified_models} are selected by the marginal likelihood criterion in $500$ trials, by sample size, for two different prior distributions.}}
    \label{tab:location_case3}
    \end{table}%
\end{example}

Finally, suppose that some of the models that we consider are correctly specified and others are misspecified in the sense of Definition \ref{def_misspecified_model}. This means that, for the latter, one or more of the active moment restrictions are invalid, or in other words, that one or more components of $V$ are incorrectly set equal to zero. Indeed, all the models for which the active moment restrictions are valid are not misspecified even if some invalid moment restrictions are included among the inactive moment restrictions. This is because there always exists a value $v\in\mathbb{R}^{d_{v_\ell}}$ that equates the invalid moment restriction. In this case, the true $v_*$ for this model will be different from the zero vector: $v_* \neq 0$ and the true value of the corresponding tilting parameter $\lambda$ will be zero.

For this situation, Theorems \ref{thm_consistency_correct_specification_2} and \ref{thm_consistency_selection_misspecification_2} together imply an interesting corollary: the marginal likelihood selects the correctly specified model that contains the maximum number of overidentifying moment conditions. Without loss of generality, denote this model by $M_1$. Then we have the following result.

\begin{cor}\label{cor_consistency_selection_misspecification}
  Let Assumptions \ref{Ass_1_NS} - \ref{Ass_3_BvM}, \eqref{Ass_identification_rate_contraction_correct_specification} and \eqref{Ass_identification_rate_contraction} be satisfied. Let us consider the comparison of $J$ different models $M_j$, $j=1,2,\ldots,J$ where $M_{1}$ satisfies Assumption \ref{Ass_0_NS} whereas $M_{j}$, $j \neq 1$ can either satisfy Assumption \ref{Ass_0_NS} or not. Then,
  \begin{displaymath}
    \lim_{n\rightarrow \infty}P\left(\log m(x_{1:n};M_{1}) > \max_{j \neq 1}\log m(x_{1:n};M_{j})\right) = 1
  \end{displaymath}
  if and only if $(p_1 + d_{v_1}) < (p_{j} + d_{v_{j}})$, $\forall j \neq 1$ such that $M_j$ satisfies Assumption \ref{Ass_0_NS}.
\end{cor}
\noindent This corollary says that, if we compare a set of models, some of them are correctly specified and the others are misspecified, then the marginal likelihood selects model $M_1$ if and only if $M_1$ is correctly specified and contains the maximum number of overidentifying moment conditions among the correctly specified models.

%=====================================================================
\section{Applications} \label{s_examples}

The techniques discussed in the previous sections have wide-ranging applications to various statistical settings, such as generalized linear models, and to many different fields,
such as biostatistics and economics. In fact, the methods discussed above can be applied to virtually any problem that, in the frequentist setting, would be approached by generalized method of moments or
estimating equation techniques. To illustrate some of the possibilities,
we consider in this section two important problems: one in the context of count regression, and the
second in the setting of instrumental variable (IV) regression.

\subsection{Count regression: variable selection}
Consider the Poisson regression model
\bee
\begin{split}\label{eq_Poisson_DGP}
y_{i}|\beta, x_{i} &\sim Poisson(\mu_{i}),\qquad i=1,\ldots,n \\
\log(\mu_{i}) &= x_{i}'\beta.
\end{split}
\eee
where $\beta=(\beta_{1}, \beta_{2}, \beta_{3})'$ and $x_{i}=(x_{1,i}, x_{2,i}, x_{3,i})'$. In this setting, suppose
we wish to learn about $\beta$ under the moment conditions
\bee \label{eq:poisson_mom}
\begin{split}
\EE\left[\left(y_{i} - \exp(\beta_{1}x_{1,i} + \beta_{2}x_{2,i} + \beta_{3}x_{3,i})\right) x_{i} \right] &= 0 \\
\EE\left[ \left(\frac{y_{i}-\exp(x_{i}'\beta)}{\sqrt{\exp(x_{i}'\beta)}}\right)^{2} -1\right] &= v.
\end{split}
\eee
The first type of moment restriction (one for each $x_{j,i}$ for $j=1,2,3$) is derived from the fact that the conditional expectation of $y_{i}$ is $\exp(x_{i}'\beta)$ and this identifies $\beta$. The second type of restriction is an overidentifying restriction that is implied by the Poisson assumption. More specifically, if $v=0$ that moment condition asserts that the conditional variance of $y_i$ is equal to the conditional mean. In general, this condition implied by the Poisson assumption can be questioned by supposing that $v \neq 0$.

Suppose that we are interested in determining if $x_3$ is a redundant regressor and if the conditional mean and variance are equal. To solve this problem, we can create the following four models based on the grand model \eqref{eq:poisson_mom} with the following restrictions:
\bee
\begin{split}\label{eq_models_Poisson_DGP}
M_{1} &: \text{$\beta_{1}$ and  $\beta_{2}$ are free parameters, $\beta_{3}=0$ and $v=0$.}\\
M_{2} &: \text{$\beta_{1}$, $\beta_{2}$, $\beta_{3}$ are free parameters and $v=0$.}\\
M_{3} &: \text{$\beta_{1}$, $\beta_{2}$ and $v$ are free parameters, and $\beta_{3}=0$.}\\
M_{4} &: \text{$\beta_{1}$ $\beta_{2}$, $\beta_3$ and $v$ are free parameters.}\\
\end{split}
\eee
As required, each model has the same moment restrictions. The different models arise from the different restrictions on $\beta_3$ and $v$. All the models are correctly specified and they only differ in terms of overidentifying moment restrictions.

Consider first the case where the data are drawn under the Poisson assumption (this information is, of course,
not used in the estimation). Specifically, suppose we generate $n$ realizations of $\{y_{i}, x_{i}\}$ from the Poisson model with $\beta_{1}=1$, $\beta_{2}=1$, $\beta_{3}=0$. Thus, $x_{3,i}$ is a redundant regressor. Each explanatory variable $x_{j,i}$ is generated $i.i.d.$ from normal distributions with mean .4 and standard deviation 1/3. Given these data, our goal is to evaluate the finite-sample performance of our marginal likelihood criterion in picking out the correct model. We conduct our MCMC analysis and compute the marginal likelihoods of the four models by the Chib (1995) method under the default student-t prior distribution on $\beta$ given in (\ref{eq:defaultprior}). The results, in Table \ref{tab:poisson_case1}, give the percentage of times in 500 replications that the marginal likelihood criterion picks each model for three different sample sizes. As can be seen, the model with the largest number of overidentifying moment restrictions $M_{1}$ is selected by the marginal likelihood criterion with frequency close to one even when $n=250$.
\begin{table}[h!]
  \centering
    \begin{tabular}{rrrrr}
  \hline
  Model & $M_{1}$ & $M_{2}$ & $M_3$ & $M_4$ \\ \hline
$n = 250$ & 0.99 & 0.01 & 0.01 & 0.00 \\
  $n = 500$ & 0.99 & 0.00 & 0.01 & 0.00 \\
  $n = 1000$ & 0.99 & 0.00 & 0.00 & 0.00 \\ \hline
    \end{tabular}%
    \caption{Frequency (\%) of times each of the four models in \eqref{eq_models_Poisson_DGP} are selected by the marginal likelihood criterion in $500$ trials. The DGP is the Poisson model given in \eqref{eq_Poisson_DGP}.}% Model $M_1$, defined by $\beta_3=0$ and $v=0$, is the true model. The other models are defined in the text. This table presents the percentage of times each model is selected by the marginal likelihood criterion in 500 trials.  The marginal likelihood is computed by the method of \cite{Chib1995} and \cite{ChibJeliazkov2001} as described in Section \ref{ss_ML}.}
    \label{tab:poisson_case1}
\end{table}

Now suppose that $y_{i}$, $i=1,\ldots,n$ are generated from the negative binomial (NB) distribution
\bee
\begin{split}
y_{i}|\beta, x_{i} &\sim NB\left( \frac{p}{1-p}\mu_{i}, \;p \right),\qquad \mu_i >0,\quad p\in (0,1) \\
\log(\mu_{i}) & = x_{i}'\beta
\end{split}
\eee
where $\mu_i$ is the size parameter. In this set-up, models $M_3$ and $M_4$ are the correctly specified models but $M_3$ has more overidentifying moment restrictions than $M_4$. For our experiment, we set $p=1/2$ and compare the performance of the marginal likelihood criterion in selecting among the four models. The results are given in Table \ref{tab:poisson_case3}. The results show that the frequency of selecting $M_3$ is 94\% for $n=250$ and this percentage increases with sample size, in accordance with our theory. In addition, neither model $M_1$ nor $M_2$ (which state equality  of the conditional mean and variance) is picked for any sample size.
\begin{table}[h!]
  \centering
    \begin{tabular}{rrrrr}
  \hline
  Model & $M_{1}$ & $M_{2}$ & $M_3$ & $M_4$ \\ \hline
$n = 250$ & 0.00 & 0.00 & 0.94 & 0.06 \\
  $n = 500$ & 0.00 & 0.00 & 0.95 & 0.05 \\
  $n = 1000$ & 0.00 & 0.00 & 0.96 & 0.04 \\ \hline
    \end{tabular}%
    \caption{Frequency (\%) of times each of the four models in \eqref{eq_models_Poisson_DGP} are selected by the marginal likelihood criterion in $500$ trials. Model choice with Negative Binomial DGP. Model $M_3$, defined by $\beta_3=0$ and $v$ free, is the true model. The other models are defined in the text.} %This table presents the percentage of times each model is selected by the marginal likelihood criterion in 500 trials. The marginal likelihood is computed by the method of \cite{Chib1995} and \cite{ChibJeliazkov2001} as described in Section \ref{ss_ML}.}
    \label{tab:poisson_case3}
\end{table}

It is important to emphasize that our analysis of the data, and the comparison across models, was light in terms of assumptions. The Poisson and negative binomial distributions were used to simply obtain a sample. These distributional forms are not featured in the estimation or the model comparison. A reader of this paper wondered how a parametric Poisson model would have performed for these data. Since the data was generated under either a Poisson model or a model close to a Poisson model, the marginal likelihood of the Poisson model (correctly) is higher than that of the moment model. But this performance suffers dramatically if the data are generated from a count process that is quite different from the Poisson. For instance, suppose that the data are generated under the assumption that the first three moment conditions hold. We have developed a way of generating such a sample which works as follows. We first generate a large population of count data from an arbitrary count process (say $y_{i} = \lfloor\exp\{\beta_{1}x_{1,i} + \beta_{2}x_{2,i} + 20 \mathcal{N}(0,1)\}\rfloor$, setting any negative observations to zero and where $\lfloor a\rfloor$ denotes the largest integer less than or equal to $a$). We then find the ETEL probabilities $p_{i}^{\ast }$ consistent with the given moment conditions. Finally, we sample the population of observations according to these probabilities. The resulting sample satisfies the moment conditions but has no connection to the Poisson or negative binomial distributional forms. For such a design, in $500$ replications, in the parametric Poisson models (one with $\beta_3=0$ and one with $\beta_3$ free), the Poisson model with $\beta_3$ = 0 is selected $42\%$ when $n = 250$, $46\%$ when $n = 500$, and $45\%$ when $n = 1000$. Thus, the Poisson assumption is not capable of selecting the correct case. In addition, when these two Poisson models are compared along with the 4 moment models in \eqref{eq_models_Poisson_DGP}, for which the marginal likelihoods are computed with our method, $M_3$ is decisively preferred over the Poisson models and the frequency of times it is selected is similar to that reported above in Table 5.

\subsection{IV Regression}

Consider now the commonly occurring situation with observational data where
one is interested in learning about the causal effect parameter $\beta $ in
the model
\begin{equation*}
y = \alpha + x\beta + w\delta + \varepsilon
\end{equation*}%
but the covariate $x$ is correlated with the error, due to say unmeasured or
uncontrolled factors, apart from $w$, that are correlated with $x$ that reside in $\varepsilon $.  Also suppose that one has two valid instrumental variables $z_{1}$ and $z_{2}$ that (by definition) are correlated with $x$ but uncorrelated with $\varepsilon$. In this setting we can learn about $\theta = (\alpha, \beta, \delta)$
from the overidentified moment restrictions
\begin{align}
\EE\left[ \left( y_{i}-\alpha -x_{i}\beta -w_{i}\delta \right) \right] & =0\label{IV_MR_1}
\\
\EE\left[ \left( y_{i}-\alpha -x_{i}\beta -w_{i}\delta \right) z_{1i}\right]\label{IV_MR_2}
& =0 \\
\EE\left[ \left( y_{i}-\alpha -x_{i}\beta -w_{i}\delta \right) z_{2i}\right]\label{IV_MR_3}
& =0 \\
\EE\left[ \left( y_{i}-\alpha -x_{i}\beta -w_{i}\delta \right) w_{i}\right]\label{IV_MR_4}
& =0, \qquad (i\leq n)
\end{align}
without having to model the distribution of $\varepsilon $ or the model
connecting $z$ to $x$.

\begin{figure}[h!]
     \centering
\begin{subfigure}[b]{1\textwidth}
        \resizebox{\linewidth}{!}{
           \input{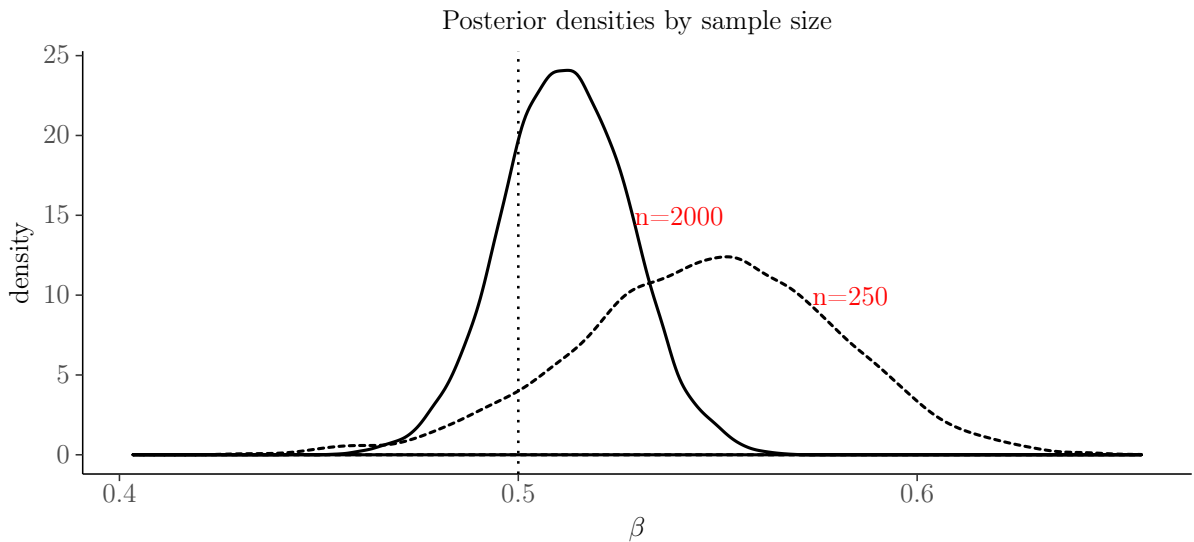}
        }
    \end{subfigure}
    \caption{\small{Posterior densities of $\beta$ in the IV regression with skewed error. Posterior densities are based on 10,000 draws beyond a burn-in of 1000. The M-H acceptance rate is about 90\% for each sample size.}}
    \label{fig:ivreg}
\end{figure}

In order to demonstrate the performance of our Bayesian prior-posterior
analysis in this setting, we generate data on $(y_{i},x_{i},z_{1i},z_{2i})$
from a design that incorporates a skewed marginal distribution of $\varepsilon $ and substantial correlation between $x$ and $\varepsilon $. In our DGP we assume that $y=1 +.5x+ .7w + \varepsilon $, $x=z_{1}+z_{2}+w + u$, and generate $z_j$ from $\mathcal{N}(.5,1)$ and $w$ from $Uniform(0,1)$. The errors $(\varepsilon,u)$ are generated from a Gaussian copula whose covariance matrix has 1 on
the diagonal, and .7 on the off-diagonal, such that the $\varepsilon $
marginal distribution is the skewed bivariate mixture $0.5\mathcal{N}(.5,.5^{2})+0.5\mathcal{N}(-.5,1.118^{2})$ and the $u$ marginal distribution
is $\mathcal{N}(0,1)$. We generate $n=250$ and $n=2000$ observations from this design and use moment conditions \eqref{IV_MR_1}-\eqref{IV_MR_4} and our default student-t prior given in (\ref{eq:defaultprior}) to learn about $\theta$.
The results shown in Table \ref{table:ivreg} and Figure \ref{fig:ivreg} demonstrate clearly the ability of our method to concentrate on the true values of the parameters,
under minimal assumptions.
\begin{table}[ht]
\centering
\begin{tabular}{rrrrrrr}
  \hline
 & mean & sd & median & lower & upper & ineff \\ \hline
  & & &$n=250$ & & & \\ \hline
$\alpha$ & 1.26 & 0.11 & 1.26 & 1.04 & 1.48 & 1.26 \\
  $\beta$ & 0.55 & 0.03 & 0.55 & 0.48 & 0.61 & 1.41 \\
  $\delta$ & 0.28 & 0.17 & 0.28 & -0.06 & 0.60 & 1.27 \\  \hline
  & & & $ n = 2000$ & & & \\ \hline
  $\alpha$ & 1.01 & 0.05 & 1.01 & 0.92 & 1.10 & 1.30 \\
  $\beta$ & 0.51 & 0.02 & 0.51 & 0.48 & 0.54 & 1.33 \\
  $\delta$ & 0.66 & 0.07 & 0.66 & 0.52 & 0.81 & 1.38 \\  \hline
\end{tabular}
\caption{\small{Posterior summary for two simulated sample sizes from IV regression model with skewed error. The true value of $\alpha$ is $1$, of $\beta$ is $.5$ and $\delta$ is $.7$. The summaries are
based on $10,000$ MCMC draws beyond a burn-in of $1000$. The M-H acceptance rate is around $90\%$ in both cases.}}
\label{table:ivreg}
\end{table}

Now suppose that we are unsure that $z_{2}$ is an appropriate instrument. We can address this concern by estimating a new model $M_2$ in which the moment condition \eqref{IV_MR_3} is not active. The marginal likelihood
of this model can be compared with the marginal likelihood of the previous model $M_1$.
The results show that for $n=250$, the log-marginal likelihood of $M_1$ is $-1395.807$ and that of $M_2$ is $-1398.092$, while for $n = 2000$, the corresponding log-marginal likelihoods are $-15217.78$ and $-15222.65$, respectively, thus correctly indicating for both sample sizes that $z_{2}$ is an appropriate instrument.

\section{Conclusion} \label{s_conclusion}

In this paper we have developed a fully Bayesian framework for estimation and model comparisons in statistical models that are defined by moment restrictions. The Bayesian analysis of such models has always been viewed as a challenge because  traditional Bayesian semiparametric methods, such as those based on Dirichlet process mixtures and variants thereof, are not suitable for such models. What we have shown in this paper is that the Exponentially Tilted Empirical Likelihood setting is an immensely useful organizing framework within which a fully Bayesian treatment of such models can be developed. We have established a number of new, powerful results surrounding the Bayesian ETEL framework including the treatment of models that are possibly misspecified. We show how the moment conditions can be reexpressed in terms of additional nuisance parameters and that the Bayesian ETEL posterior distribution satisfies a Bernstein-von Mises theorem. We have also developed a framework for comparing moment condition models based on  marginal likelihoods and Bayes factors and provided a suitable large sample theory for model selection consistency. Our results show that the marginal likelihood favors the model with the minimum number of parameters and the maximum number of valid moment restrictions. When the models are misspecified, the marginal likelihood based selection procedure selects the model that is closer to the (unknown) true data generating process in terms of the Kullback-Leibler divergence. The ideas and results illumined in this paper now provide the means for analyzing a whole array of models from the Bayesian viewpoint. This broadening of the scope of Bayesian techniques to previously intractable problems is likely to have far-reaching practical consequences.

\appendix
\section*{Appendix}
\section{Proofs for Sections \ref{s_Asymptotic_result_cs} and \ref{s_Asymptotic_result_ms}}
In this appendix we prove Theorem \ref{thm_BvM_misspecified} and Lemma \ref{lem_asymptotic_expansion_ETEL_misspecified}. Theorem \ref{thm_BvM_correctly_specified} is proved in the Supplementary Appendix. It is useful to introduce some notation that will be used in this section. The estimator $\whp := (\wt, \wh v)$ denotes \cite{Schennach2007}'ETEL estimator of $\psi$:
\begin{equation}\label{eq_ETEL_def}
  \whp := \arg\max_{\psi\in\Psi}\frac{1}{n}\sum_{i=1}^n\left[\wh\lambda(\psi)'g^A(x_i,\psi) - \log\frac{1}{n}\sum_{j=1}^n\exp\{\wh\lambda(\psi)'g^A(x_j,\psi)\}\right]
\end{equation}
\noindent where $\wh\lambda(\psi) = \arg\min_{\lambda\in\mathbb{R}^d}\frac{1}{n}\sum_{i=1}^n\left[\exp\{\lambda' g^A(x_i,\psi)\}\right]$.
The log-likelihood ratio is:
\begin{eqnarray}
  l_{n,\psi}(x) - l_{n,\psi_\circ}(x) & = & \log \frac{e^{\wh\lambda(\psi)'g^A(x,\psi)}}{\frac{1}{n}\sum_{j=1}^n\left[e^{\wh\lambda(\psi)'g^A(x_j,\psi)}\right]} - \log\frac{e^{\wh\lambda(\psi_\circ)'g^A(x,\psi_\circ)}}{\frac{1}{n}\sum_{j=1}^n\left[e^{\wh\lambda(\psi_\circ)'g^A(x_j,\psi_\circ)}\right]}.
%  & = & \lambda_n'g(x,\psi) - \lambda_n^{*'}g(x,\psi_*) - \log\E_n\left[e^{\lambda_n'g(x_j,\psi)}\right] + \log\E_n\left[e^{\lambda_n^{*'}g(x_j,\psi_*)}\right].\nonumber
\end{eqnarray}
%
%We make the following assumptions (where we denote by $B(\psi_\circ,n^{-1/2})$ a ball centred at $\psi_\circ$ with radius $n^{-1/2}$).
%\begin{assum}\label{Ass_4_misspecified_BvM}
%  \textit{(a)} $\psi_\circ\in Int(\Theta)$; \textit{(b)} $\E^P\|\partial g(X,\psi)/\partial \psi'\|^{\kappa} < \infty$ for every $\psi \in B(\psi_\circ,n^{-1/2})$ and $\kappa=2,4$; \textit{(c)} $\E^P\|g(X,\psi^{\circ})\|^{\kappa} < \infty$ for $\kappa=2,4$.
%\end{assum}
%

%% OLD Proofs for Stochastic LAN and Posterior Consistency
%==========================================================================================================================
\subsection{Proof of Theorem \ref{thm_BvM_misspecified}.}
The main steps of this proof proceed as in the proof of \cite[Theorem 10.1]{VanDerVaart2000} and \cite[Theorem 2.1]{kleijn2012} while the proofs of the technical theorems and lemmas that we use all along this proof are new. Let us consider a reparametrization of the model centred around the pseudo-true value $\psi_\circ$ and define a local parameter $h = \sqrt{n}(\psi-\psi_\circ)$. Denote by $\pi^h$ and $\pi^h(\cdot|x_{1:n})$ the prior and posterior distribution of $h$, respectively. Denote by $\Phi_n$ the normal distribution $\mathcal{N}_{\Delta_{n,\psi_\circ},V_{\psi_\circ}^{-1}}$ and by $\phi_n$ its Lebesgue density. For a compact subset $K\subset\mathbb{R}^p$ such that $\pi^h(h\in K|x_{1:n}) > 0$ define, for any Borel set $B\subseteq \Psi$,
$$\pi_{K}^h(B|x_{1:n}) := \frac{\pi^h(K\cap B|x_{1:n})}{\pi^h(K|x_{1:n})}$$
\noindent and let $\Phi_n^K$ be the $\Phi_n$ distribution conditional on $K$. The proof consists of two steps. In the first step we show that the Total Variation (TV) norm of $\pi_{K}^h(\cdot|x_{1:n})-\Phi_n^K$ converges to zero in probability. In the second step we use this result to show that the TV norm of $\pi^h(\cdot|x_{1:n})-\Phi_n$ converges to zero in probability.\\
\indent Let Assumption \ref{Ass_3_BvM} \textit{(a)} hold. For every open neighborhood $\mathcal{U}\subset \Psi$ of $\psi_{\circ}$ and a compact subset $K\subset\mathbb{R}^p$, there exists an $N$ such that for every $n\geq N$:
\begin{equation}\label{eq_BvM_1}
  \psi_{\circ} + K\frac{1}{\sqrt{n}}\subset \mathcal{U}.
\end{equation}
\noindent Define the function $f_n: K\times K \rightarrow \mathbb{R}$ as, $\forall k_1,k_2\in K$:
\begin{displaymath}
  f_n(k_1,k_2):=\left(1 - \frac{\phi_n(k_2)s_n(k_1)\pi^h(k_1)}{\phi_n(k_1)s_n(k_2)\pi^h(k_2)}\right)_+
\end{displaymath}
\noindent where $(a)_+ = \max(a,0)$, here $\pi^h$ denotes the Lebesgue density of the prior $\pi^h$ for $h$ and $s_n(h) := p(x_{1:n}|\psi_\circ + h/\sqrt{n})/p(x_{1:n}|\psi_\circ)$. The function $f_n$ is well defined for $n$ sufficiently large because of \eqref{eq_BvM_1} and Assumption \ref{Ass_3_BvM} \textit{(a)}. Remark that by \eqref{eq_BvM_1} and since the prior for $\psi$ puts enough mass on $\mathcal{U}$, then $\pi^h$ puts enough mass on $K$ and as $n\rightarrow \infty$: $\pi^h(k_1)/\pi^h(k_2)\rightarrow 1$. Because of this and by the stochastic LAN expansion \eqref{eq_stochastic_LAN} in Theorem \ref{thm_stochastic_LAN} below:
\begin{multline}
  \log\frac{\phi_n(k_2)s_n(k_1)\pi^h(k_1)}{\phi_n(k_1)s_n(k_2)\pi^h(k_2)} = -\frac{1}{2}(k_2 - \Delta_{n,\psi^{\circ}})'V_{\psi^{\circ}}(k_2 - \Delta_{n,\psi^{\circ}}) + \frac{1}{2}(k_1 - \Delta_{n,\psi^{\circ}})'V_{\psi_\circ}(k_1 - \Delta_{n,\psi_\circ})\\ \hfill
  + k_1'V_{\psi_\circ}\Delta_{n,\psi_\circ} - \frac{1}{2}k_1'V_{\psi_\circ}k_1 - k_2'V_{\psi_\circ}\Delta_{n,\psi_\circ} + \frac{1}{2}k_2'V_{\psi_\circ}k_2 + o_p(1) = o_p(1).
\end{multline}
\noindent Since, for every $n$, $f_n$ is continuous in $(k_1,k_2)$ and $K\times K$ is compact, then
\begin{equation}\label{eq_BvM_fn_converges_0}
  \sup_{k_1,k_2\in K} f_n(k_1,k_2)\overset{p}{\to} 0,\qquad \textrm{ as }n\rightarrow \infty.
\end{equation}
\indent Suppose that the subset $K$ contains a neighborhood of $0$ (which guarantees that $\Phi_n(K)>0$ and then that $\Phi_n^K$ is well defined) and let $\Xi_n := \{\pi^h(K|x_{1:n})>0\}$. Moreover, for a given $\eta>0$ define the event
%\begin{displaymath}
  $\Omega_n :=\left\{\sup_{k_1,k_2\in K}f_n(k_1,k_2)\leq \eta\right\}$.
%\end{displaymath}
\noindent The TV distance $\|\cdot \|_{TV}$ between two probability measures $P$ and $Q$, with Lebesgue densities $p$ and $q$ respectively, can be expressed as: $\|P - Q\|_{TV} = 2\int(1 - p/q)_{+}dQ$. Therefore, by the Jensen inequality and convexity of the functions $(\cdot)_+$,
\begin{multline}\label{eq_BvM_TV_conditional}
  \frac{1}{2}\E^P\|\Phi_n^K - \pi_K^h(\cdot|x_{1:n})\|_{TV}1_{\Omega_n \cap\Xi_n} = \E^P\bigintsss_{K}\left(1 - \frac{d\Phi_n^K(k_2)}{d\pi_K^h(k_2|x_{1:n})}\right)_{+}d\pi_{K}^h(k_2|x_{1:n})1_{\Omega_n \cap\Xi_n}\\\hfill
  \leq \E^P\bigintssss_{K}\bigintssss_{K} f_n(k_1,k_2)d\Phi_n^K(k_1) d\pi_{K}^h(k_2|x_{1:n})1_{\Omega_n \cap\Xi_n}\\\hfill
  \leq \E^P \sup_{k_1,k_2\in K}f_n(k_1,k_2)1_{\Omega_n \cap\Xi_n}
\end{multline}
\noindent that converges to zero by \eqref{eq_BvM_fn_converges_0}. By \eqref{eq_BvM_TV_conditional}, it follows that (by remembering that $\|\cdot\|_{TV}$ is upper bounded by $2$)
\begin{multline}\label{eq:BvM:convergence:TV:1}
  \E^P\|\pi_K^h(\cdot|x_{1:n}) - \Phi_n^K\|_{TV}1_{\Xi_n} \leq \E^P\|\pi_K^h(\cdot|x_{1:n}) - \Phi_n^K\|_{TV}1_{\Omega_n \cap\Xi_n} + 2P(\Omega_n^c \cap\Xi_n),%\\
  %& = & \E^P\|\pi_K^h(\cdot|x_{1:n}) - \Phi_n^K\|_{TV}1_{\Omega_n \cap\Xi_n} + o(1)
\end{multline}
\indent where the second term is $o(1)$ by \eqref{eq_BvM_fn_converges_0}. In the second step of the proof let $K_n$ be a sequence of closed balls in the parameter space of $h$ centred at $0$ with radii $M_n\rightarrow \infty$ and redefine $\Xi_n$ accordingly. For each $n\geq 1$, \eqref{eq:BvM:convergence:TV:1} holds for these balls. Moreover, by \eqref{eq_posterior_consistency} in Theorem \ref{thm_posterior_consistency_misspecified} below: $P(\Xi_n) \rightarrow 1$. Therefore, by the triangular inequality, the TV distance is upper bounded by
\begin{multline*}
  \E^P\|\pi^h(\cdot|x_{1:n}) - \Phi_n\|_{TV} \leq \E^P\|\pi^h(\cdot|x_{1:n}) - \Phi_n\|_{TV}1_{\Xi_n} + \E^P\|\pi^h(\cdot|x_{1:n}) - \Phi_n\|_{TV}1_{\Xi_n^c}\\
  \leq \E^P\|\pi^h(\cdot|x_{1:n}) - \pi_{K_n}^h(\cdot|x_{1:n})\|_{TV} +  \E^P\|\pi_{K_n}^h(\cdot|x_{1:n}) - \Phi_n^{K_n}\|_{TV}1_{\Xi_n}\\\hfill
   + \E^P \|\Phi_n^{K_n} - \Phi_n\|_{TV} + 2P(\Xi_n^c)\\
  \leq 2\E^P(\pi_{K_n^c}^h(\cdot|x_{1:n})) +  \E^P\|\pi_{K_n}^h(\cdot|x_{1:n}) - \Phi_n^{K_n}\|_{TV}1_{\Xi_n} + o(1)\overset{p}{\to} 0%\\
%  \leq 2\E^P(\pi_{K_n^c}^h(\cdot|x_{1:n})) +  \E^P\|\pi_{K_n}^h(\cdot|x_{1:n}) - \Phi_n^{K_n}\|_{TV}1_{\Xi_n} + 2P(\Xi_n^c) + o(1) \overset{p}{\to} 0
\end{multline*}
\noindent since $\E^P(\pi^h(K_n^c|x_{1:n})) = o(1)$ by \eqref{eq_posterior_consistency} and $\E^P\|\pi_{K_n}^h(\cdot|x_{1:n}) - \Phi_n^{K_n}\|_{TV}1_{\Xi_n} = o_P(1)$ by \eqref{eq:BvM:convergence:TV:1} and \eqref{eq_BvM_TV_conditional}, and where in the third line we have used the fact that: $\E^P\|\pi^h(\cdot|x_{1:n}) - \pi_{K_n}^h(\cdot|x_{1:n})\|_{TV} = 2\E^P(\pi^h(K_n^c|x_{1:n}))$ and $\|\Phi_n^{K_n} - \Phi_n\|_{TV} = \|\Phi_n^{K_n^c}\|_{TV} = o_p(1)$ by \cite[Lemma 5.2]{kleijn2012} since $\Delta_{n,\psi_0}$ is uniformly tight.
\begin{flushright}
  $\square$
\end{flushright}

The next theorem establishes that the misspecified model satisfies a stochastic Local Asymptotic Normality (LAN) expansion around the pseudo-true value $\psi_\circ$.

\begin{thm}[Stochastic LAN]\label{thm_stochastic_LAN}
  Assume that the matrix $V_{\psi_\circ}$ is nonsingular and that Assumptions \ref{Ass_1_NS} \textit{(a)}-\textit{(d)}, \ref{Ass_2_NS} \textit{(b)}, \ref{Ass_absolute continuity}, \ref{Ass_3_invalid_moments}, and \ref{Ass_3_BvM} hold. Then for every compact set $K\subset \mathbb{R}^p$,
\begin{equation}\label{eq_stochastic_LAN}
  \sup_{h\in K}\left|\log \frac{p(x_{1:n}|\psi_\circ + h/\sqrt{n})}{p(x_{1:n}|\psi_\circ)} - h'V_{\psi_\circ}\Delta_{n,\psi_\circ} + \frac{1}{2}h'V_{\psi_\circ}h\right|\overset{p}{\to} 0
\end{equation}
\noindent where $\psi_\circ$ is as defined in \eqref{theta_misspecified}, $V_{\psi_\circ} = - \E^P[\ddot{\mathfrak{L}}_{n,\psi_\circ}]$ and $\Delta_{n,\psi_\circ} = \frac{1}{\sqrt{n}}\sum_{i=1}^n V_{\psi_\circ}^{-1}\dot{\mathfrak{L}}_{n,\psi_\circ}(x_i)$ is bounded in probability.
\end{thm}
\proof See Supplementary Appendix. \begin{flushright}
  $\square$
\end{flushright}

The next theorem establishes that the posterior of $\psi$ concentrates and puts all its mass on $\Psi_n$ as $n\rightarrow\infty$.
\begin{thm}[Posterior Consistency]\label{thm_posterior_consistency_misspecified}
  Assume that the stochastic LAN expansion \eqref{eq_stochastic_LAN} holds for $\psi_\circ$ defined in \eqref{theta_misspecified}. Moreover, let Assumptions \ref{Ass_prior} \textit{(a)}, \ref{Ass_absolute continuity} and \ref{Ass_prior_BvM} hold and assume that there exists a constant $C>0$ such that for any sequence $M_n\rightarrow \infty$,
  \begin{equation}\label{Ass_identification_rate_contraction_appendix}
    P\left(\sup_{\psi\in \Psi_n^c}\frac{1}{n}\sum_{i=1}^n \lt(l_{n,\psi}(x_i) - l_{n,\psi_\circ}(x_i)\rt) \leq - \frac{CM_n^2}{n}\right) \rightarrow 1
  \end{equation}
  \noindent as $n\rightarrow \infty$ where $\Psi_n := \{\|\psi - \psi_\circ\| \leq M_n/\sqrt{n}\}$. Then,
  \begin{equation}\label{eq_posterior_consistency}
    \pi\left(\left.\sqrt{n}\|\psi - \psi_\circ\| > M_n\right|x_{1:n}\right)\overset{p}{\to} 0
  \end{equation}
  \noindent for any $M_n\rightarrow \infty$, as $n\rightarrow \infty$.
\end{thm}

\proof See Supplementary Appendix. \begin{flushright}
  $\square$
\end{flushright}
%==========================================================================================================================
\subsection{Proof of Lemma \ref{lem_asymptotic_expansion_ETEL_misspecified}.}
By Theorem 10 of \cite{Schennach2007}, which is valid under Assumptions \ref{Ass_1_NS} \textit{(a)}-\textit{(c)}, \ref{Ass_absolute continuity}, \ref{Ass_3_invalid_moments} \textit{(c)}, \textit{(e)} and \ref{Ass_3_BvM}: $\sqrt{n}(\wh\psi - \psi_\circ) = O_p(1)$. Denote $\wh h := \sqrt{n}(\wh\psi - \psi_\circ)$ and $\wtl h := \Delta_{n,\psi_\circ}$. Because of \eqref{eq_stochastic_LAN}, we have:
\begin{eqnarray}
  \sum_{i=1}^n \left(l_{n,\psi_\circ + \wh h/\sqrt{n}} - l_{n,\psi_\circ}\right)(x_i) & = & \frac{1}{\sqrt{n}}\sum_{i=1}^n \wh h' \dot{\mathfrak{L}}_{n,\psi_\circ}(x_i) - \frac{1}{2}\wh h'V_{\psi_\circ}\wh h + o_p(1)\label{eq1_asymptotic_expansion_ETEL_misspecified}\\
  \sum_{i=1}^n \left(l_{n,\psi_\circ + \wtl h/\sqrt{n}} - l_{n,\psi_\circ}\right)(x_i) & = & \frac{1}{2\sqrt{n}}\sum_{i=1}^n \wtl h' \dot{\mathfrak{L}}_{n,\psi_\circ}(x_i) + o_p(1)\label{eq2_asymptotic_expansion_ETEL_misspecified}.
\end{eqnarray}
By definition of $\wh \psi$ as the maximizer of $\sum_{i=1}^n l_{n,\psi}(x_i)$, the left hand side of \eqref{eq1_asymptotic_expansion_ETEL_misspecified} is not smaller than the left hand side of \eqref{eq2_asymptotic_expansion_ETEL_misspecified}. It follows that the same relation holds for the right hand sides of \eqref{eq1_asymptotic_expansion_ETEL_misspecified} and \eqref{eq2_asymptotic_expansion_ETEL_misspecified}, and by taking their difference we obtain:
\begin{equation}\label{eq3_asymptotic_expansion_ETEL_misspecified}
  -\frac{1}{2}\left(\wh h - \frac{1}{\sqrt{n}}\sum_{i=1}^n V_{\psi_\circ}^{-1} \dot{\mathfrak{L}}_{n,\psi_\circ}(x_i)\right)'V_{\psi_\circ}\left(\wh h - \frac{1}{\sqrt{n}}\sum_{i=1}^n V_{\psi_\circ}^{-1} \dot{\mathfrak{L}}_{n,\psi_\circ}(x_i)\right) + o_p(1) \geq 0.
\end{equation}
\noindent Because $-V_{\psi_\circ}$ is negative definite then $$-\frac{1}{2}\left(\wh h - \frac{1}{\sqrt{n}}\sum_{i=1}^n V_{\psi_\circ}^{-1} \dot{\mathfrak{L}}_{n,\psi_\circ}(x_i)\right)'V_{\psi_\circ}\left(\wh h - \frac{1}{\sqrt{n}}\sum_{i=1}^n V_{\psi_\circ}^{-1} \dot{\mathfrak{L}}_{n,\psi_\circ}(x_i)\right) \leq 0.$$ This and \eqref{eq3_asymptotic_expansion_ETEL_misspecified} imply that
%\begin{displaymath}
  $\left\|V_{\psi_\circ}^{-1/2} \left(\wh h - \frac{1}{\sqrt{n}}\sum_{i=1}^n V_{\psi_\circ}^{-1} \dot{\mathfrak{L}}_{n,\psi_\circ}(x_i)\right)\rt\| \overset{p}{\to} 0$
%\end{displaymath}
\noindent which in turn implies that
\begin{displaymath}
  \lt\|\left(\wh h - \frac{1}{\sqrt{n}}\sum_{i=1}^n V_{\psi_\circ}^{-1} \dot{\mathfrak{L}}_{n,\psi_\circ}(x_i)\right)\rt\| \overset{p}{\to} 0
\end{displaymath}
\noindent which establishes the result of the lemma.
\begin{flushright}
  $\square$
\end{flushright}

%==========================================================================================================================
%==========================================================================================================================
\section{Proofs for Section \ref{ss_consistency_ML}}
In this appendix we prove Theorems \ref{thm_consistency_correct_specification_2} -- \ref{thm_consistency_selection_misspecification_2} and Corollary \ref{cor_consistency_selection_misspecification}. It is useful to introduce some notation that will be used throughout this section. %For each augmented model, we make explicit the dimension of the nuisance parameter $v$ and the model in the following way: for a model $M_{\ell}$, we denote the nuisance parameter by $v_{d_\ell}$, that is, $v_{d_\ell}$ is the $(d - d_\ell)$-vector of nonzero components of $\breve v$ where $\breve v$ satisfies $f_{\ell}(\theta,\breve v) = 0$. Accordingly, we use the notation $\psi_{d\ell} := (\theta,v_{d_\ell})$ when we need to make  explicit the dimension of $v$. The estimate $\whp := (\wt, \wh v_{d_\ell})$ denotes \cite{Schennach2007}'ETEL estimator of $(\theta,v_{d_\ell})$ in model $M_{\ell}$:
Recall the notation $\psi^\ell = (\theta^\ell,v^\ell)$ and the estimator $\whp^\ell := (\wt^\ell, \wh v^\ell)$ denotes \cite{Schennach2007}'ETEL estimator of $\psi^\ell$ in model $M_{\ell}$:
\begin{equation}\label{eq_ETEL_def_model_ell}
  \whp^\ell := \arg\max_{\psi^\ell\in\Psi^\ell}\frac{1}{n}\sum_{i=1}^n\left[\wh\lambda(\psi^\ell)' g^A(x_i,\psi^\ell) - \log\frac{1}{n}\sum_{j=1}^n\exp\{\wh\lambda(\psi^\ell)'g^A(x_j,\psi^\ell)\}\right]
\end{equation}
\noindent where $\wh\lambda(\psi^\ell) = \arg\min_{\lambda\in\mathbb{R}^d}\frac{1}{n}\sum_{i=1}^n\left[\exp\{\lambda'g^A(x_i,\psi^\ell)\}\right]$. %We use the notation  $\whp_{d_\ell}$ for the ETEL estimator $\whp$ when we need to make explicit the dimension. \cite[page 659]{Schennach2007} establishes that the ETEL estimator inherits all the first-order properties of the Empirical Likelihood estimator established in \cite{NeweySmith2004}. We are going to use these properties in some of the proofs.\\
Denote $\wh g^A(\psi^\ell) := \frac{1}{n}\sum_{i=1}^n g^A(x_i,\psi^\ell)$, $\wh g_{\ell}^A:=\wh g^A(\psi^\ell)$, $$\wh L(\psi^\ell):=\exp\{\wh\lambda(\psi^\ell)'\wh g^A(\psi^\ell)\}\left[\frac{1}{n} \sum_{i=1}^n\exp\{\wh\lambda(\psi^\ell)'g^A(x_i,\psi^\ell)\}\right]^{-1}$$ and $L(\psi^\ell) = \exp\{\lambda_\circ(\psi^\ell)'\E^P[g^A(x,\psi^\ell)]\}\left(\E^P\left[\exp\{\lambda_\circ(\psi^\ell)'g^A(x,\psi^\ell)\}\right]\right)^{-1}$. Moreover, we use the notation $\Sigma_\ell = \left(\Gamma_\ell'\Delta_\ell^{-1}\Gamma_\ell\right)^{-1}$ where $\Gamma_\ell:= \E^P\left[\frac{\partial}{\partial \psi^{\ell\prime }}g^A(X,\psi_*^\ell)\right]$ $\Delta_\ell := \E^P[g^A(X,\psi_*^\ell)g^A(X,\psi_*^\ell)^{\prime }]$. %Moreover, the matrix $\Gamma = \E^P\lt[\frac{\partial g(x,\theta_*)}{\partial \theta'}\rt]$ has dimension $d\times p$ and the matrix $\Delta = \E^P\lt[g(x,\theta_*)g(x,\theta_*)'\rt]$ has dimension $d\times d$.\\
In the proofs, we omit measurability issues which can be dealt with in the usual manner by replacing probabilities with outer probabilities.\\

\subsection{Proof of Theorem \ref{thm_consistency_correct_specification_2}}
By \eqref{eq_MSC_ML} and Lemmas \ref{lem_likelihood_ET_approx} and \ref{lem_posterior_expansion} below we obtain
\begin{multline}\label{proof_consistency_eq_1_CS}
  P\left(\max_{\ell\neq j}\log m(x_{1:n};M_{\ell}) < \log m(x_{1:n};M_j)\right) = P\Big(\max_{\ell\neq j}\Big[-\frac{n}{2}\wh g_{\ell}^{A'}\Delta^{-1}\wh g_{\ell}^A + \log\pi(\whp^{\ell}|M_{\ell})\\
  - \frac{(p_{\ell} + d_{v_{\ell}})}{2}(\log n - \log (2\pi)) + \frac{1}{2}\log |\Sigma_{\ell}|\Big]
  + \frac{n}{2}\wh g_j^{A'}\Delta^{-1}\wh g_j^A + o_p(1)\\\hfill
  < \log\pi(\whp^j|M_j) - \frac{(p_j + d_{v_j})}{2}(\log n - \log (2\pi)) + \frac{1}{2}\log |\Sigma_j|\Big).
\end{multline}
Remark that $n\wh g_{j}^{A'}\Delta^{-1}\wh g_{j}^A\overset{d}{\to} \chi_{d-(p_j + d_{v_j})}^2$, $\forall j$, so that $n\wh g_{j}^{A'}\Delta^{-1} \wh g_{j}^A = O_p(1)$. Suppose first that $(p_{\ell} + d_{v_{\ell}}>p_j + d_{v_j})$, $\forall \ell \neq j$. Since $-n\wh g_{\ell}^{A'}\Delta^{-1}\wh g_{\ell}^A < 0$ for every $\ell$, we lower bound \eqref{proof_consistency_eq_1_CS} as
\begin{multline}\label{proof_consistency_eq_2_CS}
  P\left(\max_{\ell\neq j}\log m(x_{1:n};M_{\ell}) < \log m(x_{1:n};M_j)\right) \geq P\Big(\frac{n}{2} \wh g_j^{A'}\Delta^{-1} \wh g_j^A + o_p(1)\\
  < \log n \Big[\frac{\min_{\ell\neq j}(p_{\ell} + d_{v_{\ell}}) - p_j - d_{v_j}}{2} - \frac{\min_{\ell\neq j}(p_{\ell} + d_{v_{\ell}}) - p_j - d_{v_j})}{2\log n}\log(2\pi) \\\hfill
   - \frac{\log[\max_{\ell\neq j}\pi(\whp^{\ell}|M_\ell)/\pi(\whp^j|M_j)]}{\log n} - \frac{1}{2\log n}\left(\max_{\ell\neq j}\log |\Sigma_{\ell}| - \log |\Sigma_{j}|\right)\Big] \Big)\\
   = P\Big(\underbrace{\frac{n}{2} \wh g_{j}^{A'}\Delta^{-1} \wh g_{j}^A + o_p(1)}_{=:\mathcal{I}_n} < \underbrace{\log n \Big[\frac{\min_{\ell\neq j}(p_{\ell} + d_{v_{\ell}}) - p_j - d_{v_j}}{2} + \mathcal{O}_p((\log n)^{-1})\Big]}_{=:\mathcal{II}_n} \Big).
\end{multline}
Because $\mathcal{I}_n = \mathcal{O}_p(1)$ (and is asymptotically positive) and $\mathcal{II}_n$ is strictly positive as $n\rightarrow \infty$ (since $(p_{\ell} + d_{v_{\ell}}) > (p_j + d_{v_j})$, $\forall \ell \neq j$) and converges to $+\infty$, then the probability converges to $1$. This proves one direction of the statement.\\
To prove the second direction of the statement, suppose that $\lim_{n\rightarrow\infty} P(\max_{\ell\neq j}\log m(x_{1:n};M_{\ell}) < \log m(x_{1:n};M_j)) = 1$ and consider the following upper bound (which follows from \eqref{proof_consistency_eq_1_CS} and the fact that $n\wh g_{j}^{A'}\Delta^{-1} \wh g_{j}^A > 0$, $\forall n$):
\begin{multline}\label{proof_consistency_eq_3_CS}
  P\left(\max_{\ell\neq j}\log m(x_{1:n};M_{\ell}) < \log m(x_{1:n};M_j)\right)\\ \hfill
  \leq P\left(\log m(x_{1:n};M_{\ell}) < \log m(x_{1:n};M_j)\right), \qquad \forall \ell\neq j\\
   \leq P\Big(-\frac{n}{2} \wh g_{\ell}^{A'}\Delta^{-1} \wh g_{\ell}^A + o_p(1) + \log n \Big[\frac{(p_j + d_{v_j})-(p_{\ell} + d_{v_{\ell}})}{2} + \mathcal{O}_p\left(\frac{1}{\log n}\right)\Big] < 0\Big).
\end{multline}
Because the probability in the first line of \eqref{proof_consistency_eq_3_CS} converges to $1$ as $n\rightarrow\infty$ then, necessarily, the probability in the last line of \eqref{proof_consistency_eq_3_CS} converges to $1$ which is possible only if $(p_j + d_{v_j}) < (p_{\ell} + d_{v_{\ell}})$ because $\log n \left[\frac{(p_j + d_{v_j})-(p_{\ell} + d_{v_{\ell}})}{2}\right]$ is the dominating term since $-\frac{n}{2} \wh g_{\ell}^{A'}\Delta^{-1} \wh g_{\ell}^A<0$ and it remains bounded as $n \rightarrow \infty$. Since the first inequality in \eqref{proof_consistency_eq_3_CS} holds $\forall \ell\neq j$ then convergence to $1$ of the probability in the last line of  \eqref{proof_consistency_eq_3_CS} is possible only if $(p_j + d_{v_j}) < (p_{\ell} + d_{v_{\ell}})$, $\forall \ell \neq j$.
%This proves the theorem with $M_{\ell_1}$ and $M_{\ell_2}$ replaced by the notation $M_{1}$ and $M_{2}$, respectively.
  \begin{flushright}
    $\square$
  \end{flushright}

\subsection{Proof of Theorem \ref{thm_consistency_selection_misspecification_2}}
We can write $\log p(x_{1:n}|\psi^\ell;M_{\ell}) = -n\log n + n\log \wh L(\psi^\ell)$. Then, we have:
\begin{multline}\label{proof_consistency_eq_3}
  P\left(\log m(x_{1:n};M_j) > \max_{\ell\neq j}\log m(x_{1:n};M_\ell)\right) = P\Big(n\log\wh L(\psi^{j}_\circ) + \log\pi(\psi^{j}_\circ|M_{j}) - \log\pi(\psi^{j}_\circ|x_{1:n},M_{j})\\\hfill
   > \max_{\ell\neq j} [n\log\wh L(\psi^{\ell}_\circ) + \log\pi(\psi^{\ell}_\circ|M_{\ell}) - \log\pi(\psi^{\ell}_\circ|x_{1:n},M_{\ell})]\Big) \\\hfill
  = P\Big(n\log L(\psi^{j}_\circ) + n\log \frac{\wh L(\psi^{j}_\circ)}{L(\psi^{j}_\circ)} + \mathcal{B}_j > \max_{\ell\neq j}\Big[ n\log L(\psi^{\ell}_\circ) + \mathcal{B}_{\ell} + n \log \frac{\wh L(\psi^{\ell}_\circ)}{L(\psi^{\ell}_\circ)}\Big]\Big)
\end{multline}
where $\forall \ell$, $\mathcal{B}_{\ell} := \log\pi(\psi^{\ell}_\circ|M_{\ell}) - \log\pi(\psi^{\ell}_\circ|x_{1:n},M_{\ell})$ and $\mathcal{B}_{\ell} = O_p(1)$ under the assumptions of Theorem \ref{thm_BvM_misspecified}. By definition of $dQ^*(\psi)$ in Section \ref{s_Asymptotic_result_ms} we have that: $\log L(\psi^\ell_\circ) = \E^P[\log dQ^*(\psi^\ell_\circ)/dP] = -\E^P[\log dP/dQ^*(\psi^\ell_\circ)] = - K(P||Q^*(\psi^\ell_\circ))$. Remark that $\E^P[\log (dP/dQ^*(\psi_\circ^{2}))] > \E^P [\log (dP/dQ^*(\psi_\circ^{1}))]$ means that the KL divergence between $P$ and $Q^*(\psi_\circ^\ell)$, is smaller for model $M_{1}$ than for model $M_{2}$, where $Q^*(\psi_\circ^\ell)$ minimizes the KL divergence between $Q\in\mathcal{P}_{\psi_\circ^{\ell}}$ and $P$ for $\ell\in\{1,2\}$ (notice the inversion of the two probabilities).\\
First, suppose that $\min_{\ell\neq j}\E^P\left[\log \left(dP/dQ^*(\psi_\circ^{\ell})\right)\right] > \E^P\left[\log \left(dP/dQ^*(\psi_\circ^{j})\right)\right]$. By \eqref{proof_consistency_eq_3}:
\begin{multline}
  P\left(\log m(x_{1:n};M_j) > \max_{\ell\neq j}\log m(x_{1:n};M_\ell)\right) \geq \\\hfill
  P\left(\log\frac{\wh L(\psi^{j}_\circ)}{L(\psi^{j}_\circ)} - \max_{\ell\neq j}\log \frac{\wh L(\psi^{\ell}_\circ)}{L(\psi^{\ell}_\circ)}
  + \frac{1}{n}(\mathcal{B}_j - \max_{\ell\neq j}\mathcal{B}_{\ell})> \underbrace{\max_{\ell\neq j}\log L(\psi^{\ell}_\circ) - \log L(\psi^{j}_\circ)}_{=: \mathcal{I}_n}\right)
\end{multline}
This probability converges to $1$ because $\mathcal{I}_n = K(P||Q^*(\psi^j_\circ)) -  \min_{\ell\neq j}K(P||Q^*(\psi^{\ell}_\circ)) < 0$ by assumption, and $\left[\log \wh L(\psi^\ell) - \log L(\psi^\ell)\right] \overset{p}{\to}0$, for every $\psi^\ell\in\Psi^\ell$ and every $\ell\in\{1,2\}$ by Lemma \ref{Lemma_Uniform_conv_misspecified}.\\
To prove the second direction of the statement, suppose that $\lim_{n\rightarrow\infty} P(\log m(x_{1:n};M_j) > \max_{\ell\neq j}\log m(x_{1:n};M_{\ell})) = 1$. By \eqref{proof_consistency_eq_3} it holds, $\forall \ell \neq j$
\begin{multline}
  P\left(\log m(x_{1:n};M_j) > \max_{\ell\neq j}\log m(x_{1:n};M_{\ell})\right)\leq \\ \hfill
  P\Big(\log\frac{\wh L(\psi^{j}_\circ)}{L(\psi^{j}_\circ)} - \log \frac{\wh L(\psi^{\ell}_\circ)}{L(\psi^{\ell}_\circ)} + \frac{1}{n}(\mathcal{B}_j - \mathcal{B}_{\ell}) > \log \frac{L(\psi^{\ell}_\circ)}{L(\psi^{j}_\circ)}\Big).
\end{multline}
Convergence to $1$ of the left hand side implies convergence to $1$ of the right hand side which is possible only if $\log L(\psi^{\ell}_\circ) - \log L(\psi^{j}_\circ) < 0$. Since this is true for every model $\ell$, then this implies that $K(P||Q^*(\psi^j_\circ)) < \min_{\ell\neq j}K(P||Q^*(\psi^{\ell}_\circ))$ which concludes the proof.

  \begin{flushright}
    $\square$
  \end{flushright}

\subsection{Proof of Corollary \ref{cor_consistency_selection_misspecification}}
We can write $\log p(x_{1:n}|\psi^\ell;M_{\ell}) = -n\log n + n\log \wh L(\psi^\ell)$. Moreover, denote by $S_m:=\{j;\,M_j \textrm{ does not satisfy Assumption \ref{Ass_0_NS}}\}$ the set of indices of the models that are misspecified and by $S_m^c$ its complement in $\{1,2,\ldots,J\}$.\\
First, suppose that $\lim_{n\rightarrow \infty} P\left(\log m(x_{1:n};M_1) > \max_{j\neq 1}\log m(x_{1:n};M_j)\right) = 1$. Then, because $\max_{j\neq 1}\log m(x_{1:n};M_j) \geq \max_{j\neq 1; j\in S_m^c}\log m(x_{1:n};M_j)$,
\begin{multline}
  P\left(\log m(x_{1:n};M_1) > \max_{j\neq 1}\log m(x_{1:n};M_j)\right)\\
  \leq P\left(\log m(x_{1:n};M_1) > \max_{j\neq 1; \, j\in S_m^c}\log m(x_{1:n};M_j)\right)
\end{multline}
which implies that the probability on the right hand side converges to $1$ as $n\rightarrow \infty$. Then by Theorem \ref{thm_consistency_correct_specification_2}, we necessarily have $(p_1 + d_{v_1}) < (p_{j} + d_{v_{j}})$, $\forall j \neq 1$, $j\in S_m^c$.\\
%We can then use the same argument as in the second part of the proof of Theorem \ref{thm_consistency_correct_specification_2} to show that convergence to $1$ of the probability on the right hand side is possible only if $(p_1 + d_{v_1}) < (p_{j} + d_{v_{j}})$, $\forall j \neq j$.
\indent Next, suppose that $(p_1 + d_{v_1}) < (p_{j} + d_{v_{j}})$, $\forall j \neq 1$.  Define the event  $$\mathcal{A} := \left\{\max_{j\neq 1;\,j\in S_m^c}\log m(x_{1:n};M_j) > \max_{j\neq 1; j\in S_m}\log m(x_{1:n};M_j)\right\}.$$ Because all the models $M_j$ with $j\in S_m^c$ have $K(P||Q^*(\psi^j)) = 0$ (because they are correctly specified) then $\lim_{n\rightarrow 1}P(\mathcal{A}) = 1$ by Theorem \ref{thm_consistency_selection_misspecification_2}. By the Law of Total Probability we can write
\begin{multline}\label{eq_2_Cor}
  P\left(\log m(x_{1:n};M_1) > \max_{j\neq 1}\log m(x_{1:n};M_j)\right) = P\left(\left.\log m(x_{1:n};M_1) > \max_{j\neq 1}\log m(x_{1:n};M_j)\right|\mathcal{A}\right)P(\mathcal{A})\\
  + P\left(\left.\log m(x_{1:n};M_1) > \max_{j\neq 1}\log m(x_{1:n};M_j)\right|\mathcal{A}^c\right) P(\mathcal{A}^c)\\
  \geq P\left(\left.\log m(x_{1:n};M_1) > \max_{j\neq 1}\log m(x_{1:n};M_j)\right|\mathcal{A}\right) P(\mathcal{A})\\
  = P\left(\log m(x_{1:n};M_1) > \max_{j\neq 1;\,j\in S_m^c}\log m(x_{1:n};M_j)\right) P(\mathcal{A})
\end{multline}
which converges to $1$ by Theorem \ref{thm_consistency_correct_specification_2}.

  \begin{flushright}
    $\square$
  \end{flushright}

%=========================================================================================================================================================
\subsection{Technical Lemmas}
\begin{lem}\label{lem_likelihood_ET_approx}
  Let Assumptions \ref{Ass_0_NS}-\ref{Ass_2_NS} hold for $\psi^\ell$. Then,
\begin{multline}\label{eq_lem_likelihood_ET_approx}
  \log p(x_{1:n}|\whp^\ell;M_{\ell}) = -n\log n - \frac{n}{2}\wh g_{\ell}^{A'} \Delta_\ell^{-1}\wh g_{\ell}^A + o_p\lt(1\rt)\\
   = -n\log n - \frac{\chi_{d-(p_\ell + d_{v_\ell}}^2}{2} + o_p\lt(1\rt)
\end{multline}
\noindent where $\chi_{d-(p_\ell + d_{v_\ell})}^2$ denotes a chi square distribution with $(d-(p_\ell + d_{v_\ell}))$ degrees of freedom.
\end{lem}
\proof See Supplementary Appendix. \begin{flushright}
  $\square$
\end{flushright}
\begin{lem}\label{lem_posterior_expansion}
  Let Assumptions \ref{Ass_0_NS} - \ref{Ass_2_NS} and \eqref{Ass_identification_rate_contraction_correct_specification_2} hold for $\psi^\ell$. Then,
\begin{displaymath}\label{eq_BvM}
  -\log\pi(\whp^\ell|x_{1:n};M_{\ell}) = -\frac{(p_\ell + d_{v_\ell})}{2}[\log n - \log(2\pi)] + \frac{1}{2}\log |\Sigma_\ell| + o_p(1).
\end{displaymath}
\end{lem}
\proof See Supplementary Appendix. \begin{flushright}
  $\square$
\end{flushright}
\begin{lem}\label{Lemma_Uniform_conv_misspecified}
  Let $M_\ell$ be a misspecified model (that is, a model that does not satisfy Assumption \ref{Ass_0_NS}) and let $g^A(x,\psi^\ell)$ and $\psi^\ell$ be the corresponding moment functions and parameters. Then, under Assumptions \ref{Ass_1_NS} \textit{(a)}-\textit{(d)}, \ref{Ass_absolute continuity} and \ref{Ass_3_invalid_moments},
  \begin{displaymath}
    \sup_{\psi^\ell\in\Psi^\ell}\left|\log\frac{\exp\{\wh\lambda(\psi^\ell)'\wh g^A(\psi^\ell)\}}{\frac{1}{n} \sum_{i=1}^n\exp\{\wh\lambda(\psi^\ell)'g^A(x_i,\psi^\ell)\}} - \log\frac{\exp\{\lambda_\circ(\psi^\ell)'\E^P[g^A(x,\psi^\ell)]\}}{\E^P\left[\exp\{\lambda_\circ(\psi^\ell)'g^A(x,\psi^\ell)\}\right]}\right|\overset{p}{\to} 0.
  \end{displaymath}
\end{lem}
\proof See Supplementary Appendix. \begin{flushright}
  $\square$
\end{flushright} 
%\input{Technical_Lemmas_v05}

%--------------------------------------------------------------------------------------------------------------------------------------
%\clearpage
%\newpage
\mbox{}
\bibliography{AnnaBib}

\end{document}